\documentclass[sigconf, nonacm]{acmart}
\usepackage{algorithm}
\usepackage{caption}
\usepackage{graphicx}
\usepackage{subcaption}
\usepackage{balance}
\usepackage{multirow, makecell}
\usepackage[inline]{enumitem}
\usepackage{hyperref}

\usepackage{amsmath,amssymb,amsfonts}
\usepackage{algorithmic}
\usepackage{graphicx}
\usepackage{textcomp}
\usepackage{xcolor}
\usepackage{bm}
\usepackage{listings}
\usepackage[colorinlistoftodos,prependcaption,
            textsize=small, textwidth=30]{todonotes}

\usepackage{booktabs}
\usepackage{multirow}

\newcommand{\sempf}{SemPf}
\newcommand{\semplus}{Sem\textsuperscript{+}Pf}
\newcommand{\semev}{{SemEv}}
\newcommand{\nosemev}{{NoSemEv}}
\newcommand{\nosempf}{{NoSemPf}}

\newcommand\myCircled[2][]{
  \ifmmode
    \Circled[fill color=black,inner color=white,#1]{\smash{\mathsf{#2}}\vphantom{\mathsf{b}}}%
  \else
    \Circled[fill color=black,inner color=white,#1]{\smash{\textsf{#2}}\vphantom{\textsf{b}}}%
  \fi
}

\newcommand\vldbavailabilityurl{https://github.com/fzirak/semantic-dbms.git}


\begin{document}
\title{Towards Anticipatory Databases Through Shared Data and Workload Semantics}

\settopmatter{authorsperrow=4}

\author{Farzaneh Zirak}
\orcid{X}
\affiliation{%
  \institution{University of Melbourne}
}
\email{fzirak@student.unimelb.edu.au}

\author{Kasper O. Mortensen}
\orcid{X}
\affiliation{%
  \institution{Aarhus University}
}
\email{km@cs.au.dk}

\author{Farhana Choudhury}
\orcid{0000-0001-6529-4220}
\affiliation{%
  \institution{University of Melbourne}
}
\email{fchoudhury@unimelb.edu.au}

\author{Renata Borovica-Gajic}
\orcid{0000-0003-3503-4123}
\affiliation{%
  \institution{University of Melbourne}
}
\email{rborovica@unimelb.edu.au}

\begin{abstract}
    Database management systems increasingly serve dynamic and exploratory workloads, yet many of their decisions still rely on low-level signals such as recency, frequency, and address locality. These signals capture how data was accessed, but not \emph{what} is being examined or \emph{how} an analytical focus evolves. We argue for treating \emph{workload semantics} as a first-class control signal for anticipatory decision making. Central to this view, we introduce \emph{semantic locality} and \emph{semantic trajectories}, which capture relationships among nearby queries and how those relationships evolve across a session.
    
    We propose a framework that represents semantic context at the data, query, and session levels, models its evolution over time, and translates it into task-specific utility estimates. We instantiate this framework in semantic prefetching and semantic cache eviction, which share a semantic layer to make two separate decisions. Prefetching uses semantic trajectories to anticipate future accesses beyond what address-based locality can capture, while eviction uses semantic relevance to inform block replacement. These systems provide initial evidence that shared semantic context can support multiple DBMS components. We further outline how this principle can extend to other decisions and data systems, and discuss key challenges in representation, cost, adaptation, and evaluation.
\end{abstract}

\maketitle

\ifdefempty{\vldbavailabilityurl}{}{
\vspace{.3cm}
\begingroup\small\noindent\raggedright\textbf{Artifact Availability:}\\
The source code is available at \url{\vldbavailabilityurl}.
\endgroup
}

\vspace{-0.4em}\section{Introduction}

Modern database management systems (DBMSs) are increasingly workload-adaptive, using workload-, statistical-, and physical-level signals to guide decisions such as cache management~\cite{S3FIFO, ThreeLCache}, prefetching~\cite{chen2021revisiting, opdenacker2007readahead}, index tuning~\cite{dbabandit, hmab, lan2023learnedIndex, lan2024flearnedIndex, noDBA}, query optimization~\cite{qopt_neo, qopt_bao}, and system configuration~\cite{li2019qtune, li2025agenttune}. Despite their success, these signals often capture access patterns, query structure, or system state without providing insight into the \emph{content} of the accessed data and the \emph{semantics} of the queries that drive those accesses. Yet queries do not operate on arbitrary data: they target data according to particular values, properties, and relationships. We argue that these semantic relationships provide an additional source of structure for anticipating how a workload may evolve.

This opportunity is particularly apparent in data exploration, where workloads typically consist of related query sequences rather than isolated requests. Users inspect intermediate results and reformulate later queries based on what they observe~\cite{StratosExploration, noDBSIGMOD2012, idreos2015overview}. They may refine predicates, compare related data subsets, follow newly observed patterns, or shift towards nearby regions of interest. 

Consider an exploratory analysis of the Sloan Digital Sky Survey (SDSS)~\cite{abazajian2009sdss}. A user may first retrieve galaxies from a region of the sky, observe a concentration of irregular red objects, next examine objects with similar color characteristics, and then investigate their redshift distribution. Though these queries may differ in syntax and physical access, they remain related through an evolving focus in the analysis and shared characteristics of the astronomical object being investigated. This sequence also illustrates how the explored data can reveal continuity that is only apparent when successive queries are considered together.

We use \emph{semantic representation} to describe what the workload is currently examining, \emph{semantic locality} to capture how temporally nearby queries remain related through the data characteristics, attributes, relationships, or analytical interests they explore, crucially, despite syntactic or physical differences, and \emph{semantic trajectory} to describe how this context evolves over a sequence of interactions. 
Related semantic signals have been used before~\cite{tauheed2012scout, battle2016prefetching, kalinin2014interactive}, but not as a general control signal across DBMS decisions.

We envision DBMSs that treat semantic information as a \emph{first-class control signal} alongside conventional workload signals, rather than as a replacement for them. By maintaining semantic context that captures the data being examined, the attributes and relationships in focus, or how this focus evolves, the system can capture structure that recency, frequency, and resource statistics alone may miss. This context is not tied to a single representation or optimization task. Different components can combine it with established signals through task-specific utility objectives to better support future queries or data accesses that are new but semantically related.

Our prior work provides initial evidence for this direction~\cite{selep, grasp}. Using lightweight representations of data values and query accesses, we capture query trajectories to guide prefetching. In concurrent work, we apply the same principle to cache management, using semantic relationships to estimate which cached data is likely to remain useful. These systems show that semantic workload information can support multiple optimization tasks, motivating the maintenance of reusable semantic information in a shared layer rather than reconstructing it independently in each component.

In this paper, we develop this broader view through a general framework that connects \begin{enumerate*}[label=(\roman*)] \item semantic representations of data, queries, and sessions, \item semantic locality and trajectories over time, and \item task-specific future utility for DBMS decisions.\end{enumerate*} We instantiate the framework through prefetching and cache eviction, show how their semantic information can be shared, and outline a broader research agenda for semantics-driven DBMS decisions.

Concretely, this vision paper makes the following contributions:

\begin{itemize}[leftmargin=2em, topsep=2pt, itemsep=2pt, parsep=0pt]
     \item We define \emph{semantic locality} and \emph{semantic trajectories} as first-class workload signals and propose a framework to map them into task-specific utility for DBMS decisions.
    \item We instantiate and evaluate the framework for semantic prefetching and cache eviction, extending our prior systems~\cite{selep, grasp}.
    \item We outline a semantics-driven architecture with a shared semantic layer for anticipatory database systems and a research agenda for extending semantic signals across decisions and data models, addressing challenges in cost, representation, drift, and evaluation.
\end{itemize}

\section{Related Work}

    Prior work has incorporated semantic and contextual signals into a range of data-management tasks. Semantic prefetching approaches have leveraged relationships across query sequences or structural properties of explored data to predict future accesses~\cite{bowman2006semantic,tauheed2012scout,battle2016prefetching}, while interactive exploration systems have modeled regions of interest and changes in user behavior over time~\cite{kalinin2014interactive,wan2018learning}. While these studies show that semantic information can improve individual tasks, our vision treats semantic locality and trajectories as shared workload signals that can be reused across multiple DBMS components.

    A complementary body of work applies learning and adaptive techniques to individual DBMS decisions, including prefetching~\cite{chen2021revisiting}, cache replacement~\cite{LRB, ThreeLCache}, index tuning~\cite{dbabandit,hmab, indexselect_AIMeetsAI, noDBA, interpolation}, system tuning~\cite{li2019qtune,li2025agenttune}, and query optimization~\cite{qopt_bao, qopt_neo}. Such systems typically construct task-specific models from access, query, or system-state features. Our vision is orthogonal to the particular learning mechanism: a semantics-driven DBMS maintains an evolving semantic view of the workload that can complement these conventional signals and be interpreted through different utility objectives.
    
\section{Semantic Decision Framework}

\begin{figure}
   \centerline{\includegraphics[width=\linewidth]{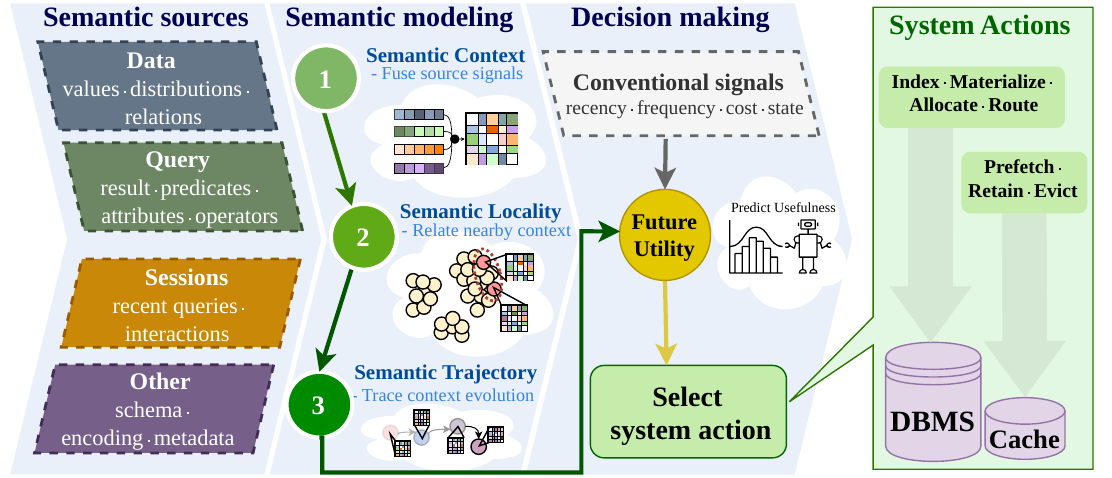}}
   \vspace{-0.3em}
   \caption{Overview of the semantics-driven decision process.}\label{fig:semantic_framework}
   \vspace{-1em}
\end{figure}

Semantic information can characterize a workload through its data, queries, and interactions. We use it in three stages, shown in Figure~\ref{fig:semantic_framework}: first, representing semantic context at an appropriate abstraction; second, relating successive states into a semantic trajectory; and finally, combining this trajectory with conventional workload and system signals to estimate future utility and guide task-specific actions. The representation, trajectory model, and decision mechanism may vary across systems and workloads while preserving this common flow from semantic context to system decisions.

\subsection{Representing Semantic Context}\label{sec:semantic_representation}

The first stage constructs a semantic representation of the current workload state from one or more levels of the workload. At the \emph{data level}, representations may capture values, distributions, entities, structural relationships, or other properties of the accessed data~\cite{battle2016prefetching, tauheed2012scout, bowman2006semantic, kalinin2014interactive}. At the \emph{query level}, they may encode predicates, attributes, operators, joins, results, or plan information~\cite{selep, grasp, dbabandit, hmab, lan2023learnedIndex, lan2024flearnedIndex}. At the \emph{session level}, information from a sequence of recent queries can be combined to characterize the current focus of an exploration.

Semantic representations at different workload levels can reinforce one another rather than being constructed independently. For example, a block representation may reflect both its contents and relationships with other blocks frequently accessed by related queries, while a query representation may combine properties of the query with semantic information derived from the data it accesses.

These representations need not explicitly recover user intent~\cite{selep, grasp, battle2016prefetching, tauheed2012scout}. Their purpose is to preserve relationships that are informative for the target system decision, while the concrete form may vary widely across systems from compact learned vectors to statistical summaries, structural features, or embeddings already maintained by the underlying system.

The appropriate representation therefore depends on available semantic information, the required object and decision granularity, and the cost of constructing and maintaining it. Richer representations may capture more expressive relationships, but incur greater storage, update, and inference cost. For online decisions, this overhead must remain compatible with the timescale of the target task.

\subsection{Semantic Locality and Trajectories}

A semantic representation captures the workload at a particular point in time, but many database decisions depend on how that state relates to recent interactions. In exploratory workloads, the sequence of representations provides information about how the analysis changes over time. Queries that access related data or examine related concepts can occupy nearby regions of the representation space, exposing the \emph{semantic locality} previously discussed.

Semantic locality becomes more informative when considered across a sequence of interactions. Rather than treating successive semantic states independently, a system can relate them over time to form a \emph{semantic trajectory}, representing how the workload moves through semantic space. A trajectory may remain within a stable semantic region as an analysis is refined, gradually move as the user follows related observations, or shift more sharply when the focus changes. Thus, the trajectory characterizes both the current workload context and its direction of change.

Such a trajectory can be constructed in various ways. A system may compare successive semantic states to capture how they change and select actions based on similarity to the current state~\cite{battle2016prefetching, tauheed2012scout, wan2018learning}. Alternatively, recent states may be modeled as a sequence to learn dependencies across contexts, or aggregated into a compact representation of the current workload~\cite{selep, grasp}.

These approaches may also consider different amounts of history: recent states often better reflect the current direction of an evolving workload, while older states can remain useful when earlier interests recur. Hence, systems may use bounded history windows, recency weighting, or adaptive mechanisms that respond to workload shifts. All these options are different realizations of the same principle that recent semantic states provide context for interpreting the workload and estimating how its focus is likely to evolve.

Viewing workloads through semantic trajectories moves semantic information beyond describing the current workload toward anticipating how it may evolve. This evolving context can then be combined with conventional workload and system signals to estimate future utility and guide task-specific decisions.

\vspace{-0.4em}
\subsection{From Semantic Context to System Actions}

Semantic context becomes actionable when it is translated into an estimate of the future utility of candidate system objects or actions. This estimate can rely on semantic information alone, or it can combine semantics with conventional workload and system signals such as recency, frequency, access cost, resource availability, or current system state. The relative importance of these signals will depend on the target decision and workload.

The notion of \emph{utility} is inherently task-specific. For prefetching, utility reflects the likelihood that a data object will be accessed soon; for cache management, it may represent whether a resident block should be retained for future reuse. Other decisions may evaluate the expected benefit of creating an index, retaining an intermediate result, allocating resources, or routing computation to a particular location. Thus, the same semantic context can inform different decisions by being interpreted through different utility~objectives.

This separation between semantic context and task-specific utility also enables multiple DBMS components to benefit from a shared view of the evolving workload. Rather than independently inferring future behavior from low-level signals, components can use common semantic context while combining it with the signals and constraints relevant to their own decisions. This provides the basis for a semantics-driven DBMS in which different system actions can be coordinated around a shared understanding of the workload.

\begin{figure}
   \centerline{\includegraphics[width=\linewidth]{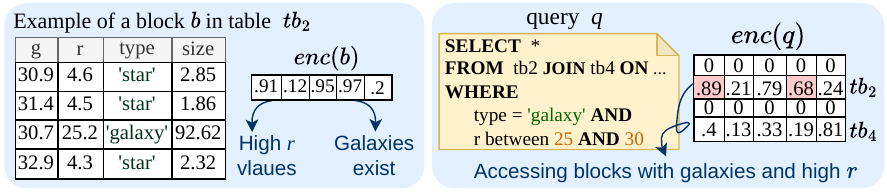}}
   \vspace{-0.3em}
   \caption{Simplified example of block and query encodings.}
   \vspace{-1.4em}
   \label{fig:enc_example}
\end{figure}
\vspace{-0.2em}
\section{Semantics in Action}

    In this section, we instantiate the framework through semantic prefetching and cache management, illustrating how shared semantic context can support distinct system decisions.

    \vspace{-0.4em}
    \subsection{Semantic Prefetching}

        Prefetching provides a first example of how semantic context can be translated into an action. Its objective is to anticipate data needed by upcoming queries and bring it into cache in advance. Conventional approaches rely on spatiotemporal locality in logical block addresses (LBAs) or recurring access patterns. However, in SQL workloads, future requests may follow relationships in the data and queries that are not apparent from the physical access stream.

        First, to construct the semantic representation, our prefetcher maps each block \(b\) from its stored values to a compact encoding \(\mathrm{enc}(b)\), capturing data characteristics independently of physical address. For a query \(q\) accessing blocks \(B_q\), block encodings are aggregated into a query-level matrix with one row per table: $\mathrm{enc}(q)=\mathrm{Agg}\left(\{\mathrm{enc}(b)\mid b\in B_q\}\right)$. 
        $\mathrm{enc}(q)$ summarizes the semantics of the accessed data and forms the state used in subsequent stages. Figure~\ref{fig:enc_example} shows examples of a block $b$, $enc(b)$, and $enc(q)$ of a query accessing it, with illustrative semantic interpretations.

\begin{figure}
           \centerline{\includegraphics[width=\linewidth]{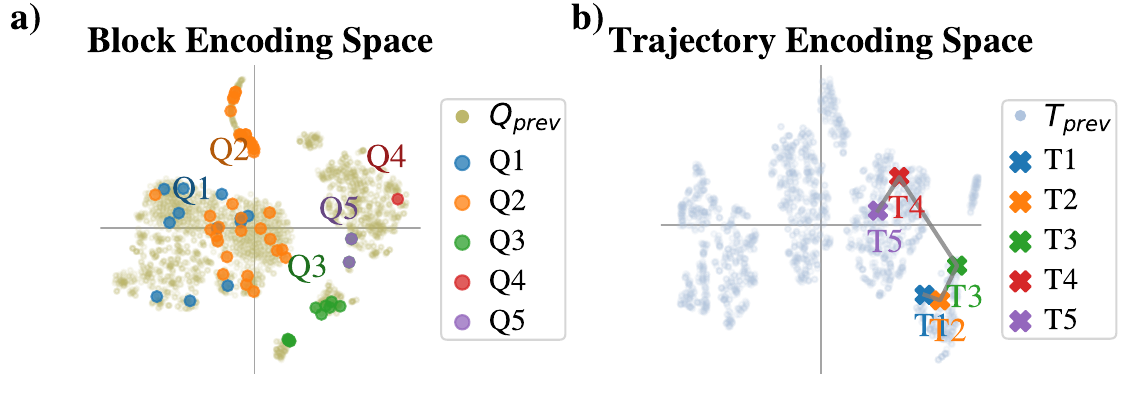}}
           \vspace{-1.3em}
           \caption{2D t-SNE plot of a) encoding of blocks accessed by 2k queries,  the last 5 highlighted, and b) their trajectories.}
           \vspace{-1.5em}
           \label{fig:case_study}
        \end{figure}

        In the second stage, our first design applies a sequence model over the \(l\) recent query representations, \(\langle enc(q_i)\rangle_{i=t-l}^{t}\), to capture dependencies across queries and derive the evolving semantic trajectory of the workload~\cite{selep}. Figure~\ref{fig:case_study}(a) shows the semantic block space for 2000 queries, with the last five highlighted. Accessed blocks form distinct semantic regions, illustrating locality in the data explored by queries. Figure~\ref{fig:case_study}(b) shows the corresponding trajectory representations, which capture how workload focus evolves over recent queries; for example, the shift from Q3 to Q4. 
        
        In the third stage, the learned semantic trajectory is used by a classification model to estimate the likelihood of future data accesses. Next, the estimates translate into prefetching decisions by selecting the data with the highest predicted utility. Experiments across SQL-based and navigational workloads showed substantial gains over address-based prefetchers (details in \S\,\ref{sec:experiments}), showing~that semantic information alone can effectively predict future accesses.

        A second design follows the same three-stage process while combining semantic and address-based information~\cite{grasp}. In the first stage, it retains the same data-derived block encodings and aggregates them into query-result representations, while enriching the semantic context with query characteristics such as accessed tables and predicate embeddings. These representations are then modeled sequentially to capture the semantic trajectory, which is combined with patterns in LBA changes across queries. This hybrid formulation preserves information about the accessed data and query structure while exploiting reusable address-space patterns, supporting prediction across larger, evolving, and transactional datasets. The resulting prefetcher remains effective across analytical and transactional workloads and generalizes to substantially larger and shifted datasets with lightweight adaptation.
        
        Together, these designs illustrate two roles for semantic information in system decision making within our framework: it can directly drive a decision when semantic relationships provide sufficient predictive structure, or complement conventional address-based signals when additional system context is beneficial. 

        \begin{figure*}[t]
                \centering
                \begin{minipage}[t]{0.49\textwidth}
                    \centering
                    \includegraphics[width=\linewidth]{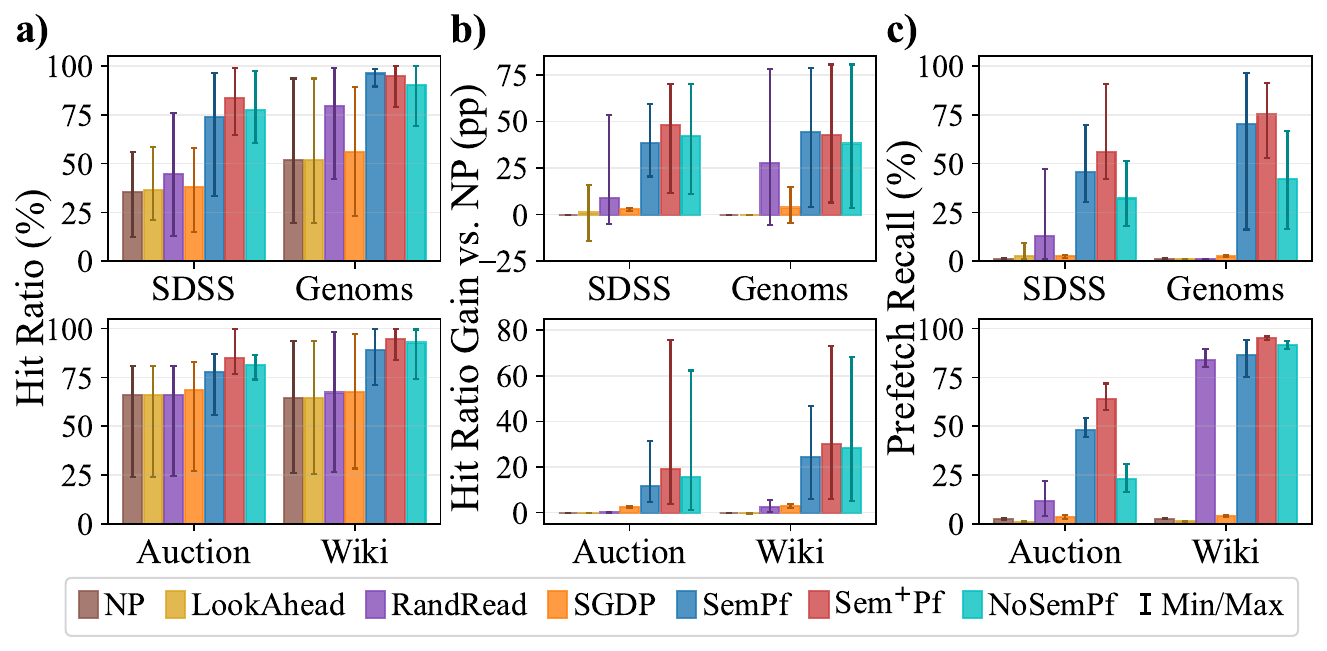}
                    \vspace{-2.3em}
                    \captionof{figure}{Prefetching performance with and without semantic context across analytical (top) and HTAP (bottom) datasets.}
                    \label{fig:prefetch_result}
                \end{minipage}
                \hfill
                \begin{minipage}[t]{0.49\textwidth}
                    \centering
                    \includegraphics[width=\linewidth]{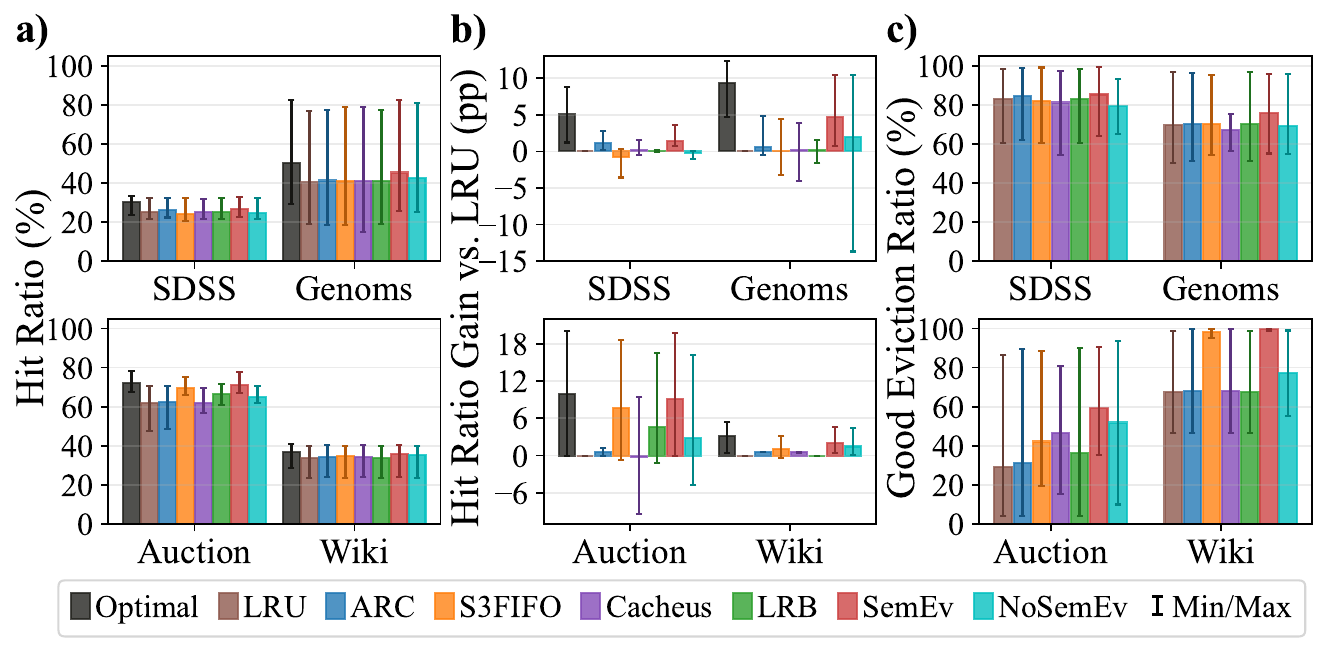}
                    \vspace{-2.3em}
                    \captionof{figure}{Cache-eviction performance with and without semantic context across analytical (top) and HTAP (bottom) datasets.}
                    \label{fig:eviction_result}
                \end{minipage}
            \vspace{-1.1em}
        \end{figure*}
    
\vspace{-0.3em}\subsection{Semantic Cache Eviction}
    
        Cache management provides a complementary use of semantic context. When a new block must be inserted into a full cache $\mathcal{C}$, the system must determine which resident block is least likely to remain useful and can be evicted. With complete future knowledge, an optimal policy evicts the block whose next access occurs farthest in the future. Since future accesses are unknown, practical policies approximate this objective using signals available at decision time, such as recency, frequency, and reuse history. Our goal is to ad- ditionally consider whether a cached block remains semantically relevant to the workload currently unfolding.
        
        Within our framework, the first stage again constructs semantic context from data and queries, reusing the semantic representations introduced for prefetching. To capture the evolving workload in the second stage, recent query representations are further aggregated into a session encoding, with greater weight assigned to more recent queries. This recency-weighted representation summarizes the current semantic trajectory and can be used to evaluate cached blocks and assess their relevance to the ongoing workload.
        
        This semantic relevance provides only one part of the information needed for eviction. At each eviction point, the context of a candidate block combines its semantic relevance with the block's conventional access signals, the current query, and the cache state. The resulting context for a candidate block $b$ can be represented as
        \begin{equation}
            x_b =
            [\phi_{\mathrm{sem}}(b,S),
            \mathrm{rec}(b),
            \mathrm{freq}(b),
            \mathrm{rd}(b),
            x_q,
            x_{\mathcal{C}}],
        \end{equation}
        where $\phi_{\mathrm{sem}}(b,S)$ measures the relationship between block $b$ and the recent session context $S$, $\mathrm{rec}$, $\mathrm{freq}$, and $\mathrm{rd}$ capture its access history, and $x_q$ and $x_{\mathcal{C}}$ summarize the current query and cache state. In our implementation, $\phi_{\mathrm{sem}}(b,S)$ is computed through an element-wise interaction between the encoding of block $b$ and the session representation maintained for its table.
        
        Consequently, the semantic component of $x_b$ reflects the second stage of the framework by situating each candidate block within the current semantic trajectory (via $\phi_{\mathrm{sem}}$ and $S$), while the complete context combines this information with conventional signals for the subsequent eviction decision.
        
        The third stage maps this context into a task-specific utility estimate for eviction. Evaluating every cached block with a learned model would be prohibitively expensive. Hence, we sample a small set of eviction candidates and use a contextual bandit~\cite{qin2014contextual} to estimate their eviction utility and select a victim. The bandit adapts online through delayed feedback: evicting a block that is reused shortly afterward receives a negative reward, while blocks that remain unused for longer receive increasingly positive feedback.

    \vspace{-0.3em}\subsection{Experimental Evidence}\label{sec:experiments}

        \subsubsection{Setup}
            {\textbf{Baselines.}} We evaluate the semantic-only prefetcher (\sempf) and the semantic+address prefetcher (\semplus) against conventional and state-of-the-art learned address-based approaches, LookAhead, random readahead (RandRead), SGDP~\cite{smith1978lookahead, opdenacker2007readahead, 2023sgdp}, and the no-prefetching system (NP). We compare the semantic eviction policy (\semev) against LRU, ARC, and S3FIFO~\cite{LRU, lru3_arc, S3FIFO}, and the learned policies Cacheus and LRB~\cite{cacheus, LRB}, with the optimal policy as a theoretical upper bound.
        
        \textbf{Datasets.} We use 90\,GB SDSS~\cite{abazajian2009sdss} and 10\,GB Genomes~\cite{sqlshare} for analytical workloads, and 16\,GB Auction and 21\,GB Wiki~\cite{DifallahPCC13} for HTAP. We evaluate 6–8 workloads per dataset, with 1000, 450, 400, and 600 queries for SDSS, Genomes, Auction, and Wiki, respectively.

        \textbf{Metrics.} We report hit ratio, i.e., the fraction of accesses served from cache, and hit-ratio gain relative to NP for prefetchers and LRU for eviction. Prefetch recall measures the fraction of future accesses identified, while good-eviction ratio quantifies eviction quality relative to the optimal policy.

        \subsubsection{Effectiveness across decisions}

            Figures~\ref{fig:prefetch_result} and~\ref{fig:eviction_result} show the performance of the prefetchers and eviction policies, respectively. For each dataset, bars report the average across its test workloads, while error bars show the observed minimum and maximum, capturing variation across workloads with different characteristics.

            \textbf{Prefetching.}
            Figure~\ref {fig:prefetch_result}(a) shows a clear hit ratio advantage for the semantic prefetchers across datasets. Averaged over all test workloads, \semplus{} reaches an 88.91\% hit ratio and \sempf{} 84.35\%, well above RandRead at 64.25\%, the strongest baseline on average. The gains over NP in Figure~\ref{fig:prefetch_result}(b) are consistently positive for both semantic approaches, while RandRead and LookAhead occasionally degrade performance by prefetching irrelevant blocks that displace useful data. These improvements also translate into runtime gains: \sempf{} reduces I/O time by up to 84\%, while \semplus{} achieves up to 90.8\% lower I/O time and 57.1\% lower end-to-end latency.

            The benefit of semantics is even clearer in the quality of the prefetch decisions. As shown in Figure~\ref{fig:prefetch_result}(c), most baselines achieve low prefetch recall, indicating that their hit-ratio gains often come from previously prefetched blocks that remain in the cache rather than accurate prediction of the immediate future accesses. In contrast, \sempf{} and \semplus{} achieve substantially higher recall across both analytical and HTAP workloads, reaching up to 94.1\%. Their stronger minimum values further show that this advantage remains consistent across workloads with different characteristics.

            Comparing \sempf{} and \semplus{} shows that both perform strongly on analytical workloads with static underlying data. However, in HTAP workloads, combining semantic and address-based signals allows \semplus{} to outperform \sempf{}, especially on Auction, which has a higher proportion of transactional queries.

            \textbf{Cache eviction.}
            Figure~\ref{fig:eviction_result}(a) shows that \semev{} achieves the strongest and most stable performance across analytical and HTAP workloads. Figure~\ref{fig:eviction_result}(b) further shows consistent gains over LRU, reaching 19.28\% with a 5.36\% average improvement, while several baselines show negative gains on some workloads. S3FIFO is the strongest baseline on average but varies substantially and falls below LRU in several cases. Cacheus and LRB are competitive on selected workloads but are less consistent across datasets.
            
            The same pattern appears in the eviction quality shown in Figure~\ref{fig:eviction_result}(c). Good eviction ratio (GER) measures the fraction of eviction decisions that match the optimal choice at that cache state~\cite{LRB}. \semev{} achieves a high GER across all datasets, with 79.5\% of its evictions matching the optimal decision on average. In contrast, the competitive baselines show greater variation and less consistent performance across workloads, confirming that semantic context can improve the robustness and quality of eviction decisions.

        \subsubsection{Reusing the semantic context} 
            Our semantic models share the same underlying semantic representations, and the results in Figures~\ref{fig:prefetch_result} and~\ref{fig:eviction_result} show that these representations can support different decision models. To isolate the contribution of semantic context, we additionally evaluate \semplus{} and \semev{} with their semantic inputs removed, denoted as \nosempf{} and \nosemev{}, respectively.

            The ablation results confirm that the semantic context is a substantial contributor to both decisions. For prefetching, removing semantics reduces hit ratio by 4.13--32.18\% and lowers prefetch recall by 25.80\% on average, with a maximum drop of 45.64\%. Thus, while \nosempf{} can exploit the remaining workload signals, those signals do not recover the predictive value provided by semantics.

            The effect is even stronger for cache eviction. Relative to \semev{}, \nosemev{} loses up to 16.47\% in hit ratio, occasionally falls below LRU, and its GER drops by up to 44.12\%. Overall, these results show that the shared semantic representations provide useful information to both prefetching and eviction, despite the two components using it through different decision models and utility objectives.

        \subsubsection{Overhead}

            Table~\ref{tab:times} reports the two main costs of using semantics: the one-off construction of reusable block representations and the per-decision generation of semantic context and trajectories. In our implementation, these representations are produced through block encoding, which can be performed offline or incrementally as new blocks are accessed; in the latter case, 100 blocks can be en- coded in under a minute. In contrast, context generation lies on the critical decision path and must remain lightweight. The prefetchers construct trajectories by applying a learning model over recent query representations, while \semev{} uses a simpler aggregation- based representation. In all cases, this cost remains low: below 7\,ms per prefetch decision and 0.5\,ms per eviction decision, supporting the practicality of using semantic context online.

            \begin{table}[htbp]
            \caption{Block representation and context generation time}
        \scalebox{0.8}{
            \setlength{\tabcolsep}{3pt}
            \setlength{\tabcolsep}{3pt}
            \renewcommand{\arraystretch}{1.25}
            \begin{tabular}{l|lrrrr}
            \toprule
            \addlinespace[-0.1em]
            \multicolumn{2}{l}{\textbf{Operation}}
            & \textbf{SDSS} & \textbf{\,Genoms} & \textbf{Wiki} & \textbf{Auction} \\
    
            \cmidrule(l{-1pt}r{1pt}){1-2}
            \cmidrule(lr{1pt}){3-3}
            \cmidrule(lr{1pt}){4-4}
            \cmidrule(lr{1pt}){5-5}
            \cmidrule(lr{1pt}){6-6}

            \addlinespace[-0.1em]
            \multirow{2}{*}{%
                \rotatebox[origin=c]{90}{\small\textbf{One-off}}}
            & Block representation\,:\,100 blocks (s) & 20.71 & 3.37 & 3.6 & 59.17 \\ 
            & Block representation\,:\,full DB (s) & 2898.17 & 1997.86 & 1665.41 & 4307.94 \\ 
    
            \addlinespace[0.25em]
            \midrule
            \addlinespace[0.2em]
    
            \multirow{3}{*}{%
                \rotatebox[origin=c]{90}{\small\textbf{Per-decision}}}
            & \sempf{} context generation (ms) & 1.98 & 4.12 & 4.76 & 2.91 \\ 
            & \semplus{} context generation (ms) & 3.06 & 6.31 & 6.57 & 4.57 \\ 
            & \semev{} context generation (ms) & 0.2591 & 0.3407 & 0.3481 & 0.459 \\ 
            
            \addlinespace[0.5em]
            \bottomrule
            \end{tabular}
            \label{tab:times}
        }
    \end{table}   

\vspace{-0.3em}\section{Toward a Semantics-Driven DBMS}

    The prefetching and cache-eviction examples motivate a broader architecture in which semantic information is maintained as a shared system resource rather than reconstructed for each task. In a \emph{semantics-driven} DBMS, a \emph{semantic control plane} derives and maintains semantic representations from both the data and users' interactions with the system. The control plane accesses relevant data, query activity, and execution information, while maintaining the auxiliary state needed to construct consistent representations, such as learned transformations or embeddings, normalization statistics, value ranges, and preprocessing metadata. As queries execute, it updates data-, query-, and session-level representations and tracks workload evolution either per user under concurrent access or across interactions. These representations then serve as shared semantic building blocks that different DBMS components can compose into the context required for their decisions.

    \begin{figure}
           \centerline{\includegraphics[width=0.98\linewidth]{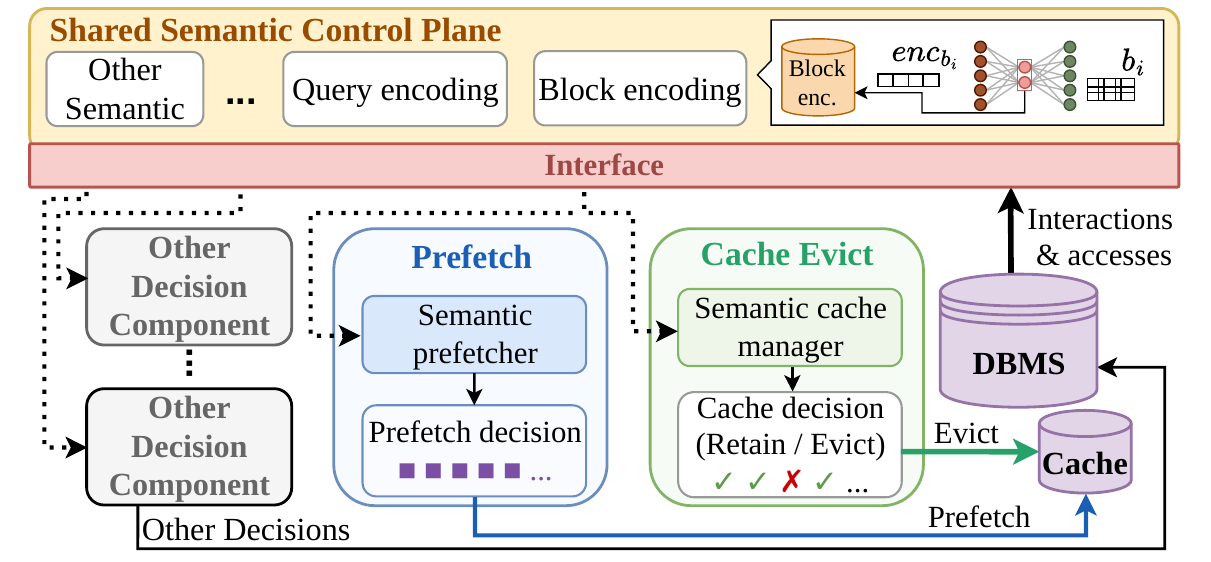}}
           \vspace{-0.3em}
           \caption{System architecture: a shared semantic layer supports semantic prefetching and semantic cache eviction.}
           \vspace{-1.5em}
           \label{fig:semantic_physical}
        \end{figure}
    
    These representations can be exposed through an \emph{interface} that supports different granularities and forms of semantic context. A component may request representations of different levels, and may consume either a sequence of recent states or an aggregation of them. Components can retrieve and compose this shared information according to their own needs, while combining it with task-specific signals and decision logic. The control plane therefore need not impose a single representation or model, but instead provides reusable semantic building blocks that support different notions of utility across the system. Figure~\ref{fig:semantic_physical} shows this architecture.

    The opportunities for semantic context extend well beyond prefetching and cache eviction. By capturing how workload focus evolves, a semantics-driven DBMS can anticipate not only which data may be accessed next, but also what kinds of queries, operations, or resources are likely to become relevant.

    For \emph{index tuning} and \emph{materialization}, recurring semantic trajectories may indicate attributes, data subsets, or intermediate results likely to become useful in upcoming queries~\cite{hmab, dbabandit, indexselect_AIMeetsAI, lan2023learnedIndex, lan2024flearnedIndex}. Such information could complement existing cost and utility estimates when deciding whether to create or retain an index, materialized view, or intermediate result. Semantic context could also complement conventional signals in \emph{resource management}~\cite{li2025agenttune, li2019qtune, huang2024sibyl, 2016clay}. Predicting the likely characteristics of the upcoming workload activity can help the system allocate memory, CPU, or I/O resources more effectively. In distributed settings, semantic information could also inform \emph{routing} and placement decisions by identifying related data or computations likely to be accessed together~\cite{smartRouting, kumar2014sword, curino2010schism}.

    Semantic context also opens opportunities for more interactive forms of assistance. A system could recommend semantically related data objects, regions, or entities for further exploration, or suggest and transform queries toward related parts of the data~space \cite{ResearcherGuide, exploract, dimitriadou2014explore,dimitriadou2016aide, workloadQueryRecom,aufaure2013recom4, drosou2013ymaldb, ma2021metainsight}. Such \emph{data morphing} and \emph{query recommendation} could turn semantic trajectories from an internal optimization signal into a mechanism for actively guiding exploration.

\vspace{-0.3em}\section{Beyond Relational Systems}

    The semantics-driven framework is not limited to relational data or SQL workloads. Its principles can extend to other data systems, with semantic context and trajectories shaped by the structure of the data and the interactions they support. A direct example is \emph{specialized engines for interactive visual exploration}~\cite{battle2013dynamic}, where users navigate through views or multidimensional regions, either locally or across distant regions that remain related through the data being explored. Therefore, tile-, cube-, and array-oriented engines provide a promising setting for semantics-driven decisions that anticipate where an exploration is likely to move next~\cite{battle2013dynamic, doshi2003adaptive, kalinin2014interactive, tauheed2012scout}.
        
    \emph{Vector databases} offer strong opportunities for this approach since semantic relationships are already represented through embeddings and vector similarity~\cite{vectorDBsearch, goccer2026qvcache}. Search sequences can form trajectories through this space, creating anticipatory signals for prefetching vectors or index regions, or allocating resources to portions of the index likely to be searched next. Importantly, the semantic control plane can reuse representations already maintained by the vector system rather than constructing them independently.

    \emph{Graph databases} \cite{erling2012virtuoso, fernandes2018graph} provide another promising setting, as semantic meaning can complement the graph's structural relationships. 
    For instance, knowledge graphs may be extended with a semantic layer that augments entities, relations, and triples with semantic annotations maintained separately from the structural elements~\cite{bainson2026sankofa}. 
    Such annotations could capture context-dependent meanings, associations, or allusions that are relevant to a particular user or analytical session but not represented in a global graph.
   
    This separation would let a graph DBMS reason jointly about the subgraph being traversed and the context-dependent associations. Semantic context could combine the currently accessed subgraph and relevant annotations; semantic locality could relate entities or subgraphs even when they are structurally distant; and semantic trajectories could capture how the workload moves through these structural and semantic spaces.
    Such signals open opportunities to anticipate shifts toward related concepts, prefetch relevant subgraphs before queries reach them, and proactively direct computation toward likely regions of exploration.
    The introduction of a unifying \emph{Graph Query Language Standard} (GQL)~\cite{gql_standard} has renewed interest in rethinking and adapting established relational design principles for graph systems~\cite{egger2026structural, katjaEDBT27aNorm}. This momentum creates a timely opportunity to consider semantics-driven mechanisms in the continuing development of graph DBMS design.  
    
\vspace{-0.3em}\section{Research Challenges and Agenda}

    \paragraph{Cost.}
        Maintaining and utilizing semantic context introduces computation, storage, and update overhead, which can become significant at large scale even when individual representations are compact. However, exploratory workloads often access only a small fraction of the available data~\cite{ResearcherGuide, StratosExploration}, creating opportunities for selective and incremental representation management. A semantics-driven DBMS could construct representations as new regions are explored, retain only those associated with active data in fast memory, and share them across components to amortize construction~cost.

        Online overhead also depends on how often semantic context must be updated and consumed. Different decisions may tolerate periodically refreshed context or require more immediate updates, creating a design space between representation richness, freshness, and decision benefit. Promising directions include employing techniques such as offline preprocessing~\cite{battle2016prefetching, selep, grasp, chatzopoulou2009recom2, wu2023factorjoin}, selective recomputation~\cite{mortensen2026practical, sato2025fast}, lightweight representations~\cite{dbabandit, li2019qtune, Learnedcardinality}, and adaptive update policies ~\cite{olma2017slalom, dbabandit, hmab, noDBA} which could inspire mechanisms that trigger or scale semantic updates only when needed.   
            
    \paragraph{Representation and granularity.}
        Semantic context can be formed at different levels and with varying richness\,\S\,\ref{sec:semantic_representation}, creating a design space around what should be shared and what should remain task-specific. A useful representation must preserve information relevant to the target decision without introducing unnecessary cost, while a shared semantic layer should support components with different granularities and objectives. This raises opportunities to study when a common representation is sufficient, when task-specific refinement is useful, and whether representation granularity can adapt dynamically to the workload and decision needs~\cite{du2023multi, dyn_repr_adapt, onlineTune, indexAdvisor}.

    \paragraph{Drift and adaptation.}
        Semantic relationships and workload trajectories change as users shift their interests, access patterns evolve, or data is inserted or updated~\cite{drift, dsb_drift}. This creates opportunities for semantic-based DBMSs to adapt both their representations and decision models over time. This raises questions about when~existing representations remain valid, when and how they should be incrementally updated or recomputed, and how quickly decision models should respond to these changes~\cite{onlineTune, grasp, Halp, mortensen2026practical, hmab, auer2019adaptively}. The goal is to preserve transferability across related workloads while remaining responsive to shifts in both semantic content and~workload~focus.
        
    \paragraph{Evaluation and benchmarks.}
        Evaluating semantics-driven decisions should go beyond prediction accuracy to capture the quality and system-level impact of the resulting actions. Depending on the component, relevant measures may include hit ratio, I/O reduction, latency, throughput, or resource use, together with the cost of constructing and maintaining semantic context. A broader evaluation agenda should also test robustness across workload shifts, datasets, and scales, and develop benchmarks that reveal when semantic signals provide consistent benefits beyond the conditions in which the representations and decision models were developed~\cite{drift, dsb_drift, shifting_workload}.

\vspace{-0.3em}
\section{Conclusion}
    In this paper, we propose treating semantic context and its evolution as first-class signals for anticipatory DBMS decisions. Through semantic prefetching and cache eviction, we demonstrate how shared semantic representations can support distinct components alongside conventional workload signals. We further outline a semantics-driven architecture built around a shared semantic control plane and discuss extensions to additional decisions and data systems as well as the existing challenges. This vision points toward DBMSs that follow the semantic direction of a workload and prepare data, computation, and resources for where it is likely to go next.

\balance
\bibliographystyle{ACM-Reference-Format}
\bibliography{refs}


\begin{thebibliography}{73}


\ifx \showCODEN    \undefined \def \showCODEN     #1{\unskip}     \fi
\ifx \showDOI      \undefined \def \showDOI       #1{#1}\fi
\ifx \showISBNx    \undefined \def \showISBNx     #1{\unskip}     \fi
\ifx \showISBNxiii \undefined \def \showISBNxiii  #1{\unskip}     \fi
\ifx \showISSN     \undefined \def \showISSN      #1{\unskip}     \fi
\ifx \showLCCN     \undefined \def \showLCCN      #1{\unskip}     \fi
\ifx \shownote     \undefined \def \shownote      #1{#1}          \fi
\ifx \showarticletitle \undefined \def \showarticletitle #1{#1}   \fi
\ifx \showURL      \undefined \def \showURL       {\relax}        \fi
\providecommand\bibfield[2]{#2}
\providecommand\bibinfo[2]{#2}
\providecommand\natexlab[1]{#1}
\providecommand\showeprint[2][]{arXiv:#2}

\bibitem[\protect\citeauthoryear{??}{gql}{2024}]%
        {gql_standard}
 \bibinfo{year}{2024}\natexlab{}.
\newblock \bibinfo{title}{Information technology -- Database languages -- GQL}.
\newblock
\newblock
\urldef\tempurl%
\url{https://www.iso.org/standard/76120.html}
\showURL{%
\tempurl}


\bibitem[\protect\citeauthoryear{Abazajian, Adelman-McCarthy, Ag{\"u}eros,
  et~al\mbox{.}}{Abazajian et~al\mbox{.}}{2009}]%
        {abazajian2009sdss}
\bibfield{author}{\bibinfo{person}{Kevork~N Abazajian},
  \bibinfo{person}{Jennifer~K Adelman-McCarthy}, \bibinfo{person}{Ag{\"u}eros},
  {et~al\mbox{.}}} \bibinfo{year}{2009}\natexlab{}.
\newblock \showarticletitle{The seventh data release of the Sloan Digital Sky
  Survey}.
\newblock \bibinfo{journal}{\emph{The Astrophysical Journal Supplement Series}}
  \bibinfo{volume}{182}, \bibinfo{number}{2} (\bibinfo{year}{2009}),
  \bibinfo{pages}{543}.
\newblock


\bibitem[\protect\citeauthoryear{Alagiannis, Borovica, Branco, Idreos, and
  Ailamaki}{Alagiannis et~al\mbox{.}}{2012}]%
        {noDBSIGMOD2012}
\bibfield{author}{\bibinfo{person}{Ioannis Alagiannis}, \bibinfo{person}{Renata
  Borovica}, \bibinfo{person}{Miguel Branco}, \bibinfo{person}{Stratos Idreos},
  {and} \bibinfo{person}{Anastasia Ailamaki}.} \bibinfo{year}{2012}\natexlab{}.
\newblock \showarticletitle{{NoDB}: Efficient Query Execution on Raw Data
  Files}. In \bibinfo{booktitle}{\emph{SIGMOD}}. \bibinfo{pages}{241--252}.
\newblock


\bibitem[\protect\citeauthoryear{Auer, Gajane, and Ortner}{Auer
  et~al\mbox{.}}{2019}]%
        {auer2019adaptively}
\bibfield{author}{\bibinfo{person}{Peter Auer}, \bibinfo{person}{Pratik
  Gajane}, {and} \bibinfo{person}{Ronald Ortner}.}
  \bibinfo{year}{2019}\natexlab{}.
\newblock \showarticletitle{Adaptively tracking the best bandit arm with an
  unknown number of distribution changes}. In
  \bibinfo{booktitle}{\emph{Conference on learning theory}}. PMLR,
  \bibinfo{pages}{138--158}.
\newblock


\bibitem[\protect\citeauthoryear{Aufaure, Kuchmann-Beauger, Marcel, Rizzi, and
  Vanrompay}{Aufaure et~al\mbox{.}}{2013}]%
        {aufaure2013recom4}
\bibfield{author}{\bibinfo{person}{Marie-Aude Aufaure},
  \bibinfo{person}{Nicolas Kuchmann-Beauger}, \bibinfo{person}{Patrick Marcel},
  \bibinfo{person}{Stefano Rizzi}, {and} \bibinfo{person}{Yves Vanrompay}.}
  \bibinfo{year}{2013}\natexlab{}.
\newblock \showarticletitle{Predicting your next OLAP query based on recent
  analytical sessions}. In \bibinfo{booktitle}{\emph{International Conference
  on Data Warehousing and Knowledge Discovery}}. Springer,
  \bibinfo{pages}{134--145}.
\newblock


\bibitem[\protect\citeauthoryear{Bainson, Mortensen, Zaporojets, Mottin, and
  Karras}{Bainson et~al\mbox{.}}{2026}]%
        {bainson2026sankofa}
\bibfield{author}{\bibinfo{person}{Ama~Bembua Bainson},
  \bibinfo{person}{Kasper~Overgaard Mortensen}, \bibinfo{person}{Klim
  Zaporojets}, \bibinfo{person}{Davide Mottin}, {and}
  \bibinfo{person}{Panagiotis Karras}.} \bibinfo{year}{2026}\natexlab{}.
\newblock \showarticletitle{Sankofa: Online Query-adaptive Dynamic Graph
  Summaries}.
\newblock \bibinfo{journal}{\emph{Proc. {VLDB} Endow.}} \bibinfo{volume}{19},
  \bibinfo{number}{10} (\bibinfo{year}{2026}), \bibinfo{pages}{2908--2921}.
\newblock
\urldef\tempurl%
\url{https://doi.org/10.14778/3828612.3828641}
\showDOI{\tempurl}


\bibitem[\protect\citeauthoryear{Battle, Chang, and Stonebraker}{Battle
  et~al\mbox{.}}{2016}]%
        {battle2016prefetching}
\bibfield{author}{\bibinfo{person}{Leilani Battle}, \bibinfo{person}{Remco
  Chang}, {and} \bibinfo{person}{Michael Stonebraker}.}
  \bibinfo{year}{2016}\natexlab{}.
\newblock \showarticletitle{Dynamic Prefetching of Data Tiles for Interactive
  Visualization}. In \bibinfo{booktitle}{\emph{Proceedings of the International
  Conference on Management of Data, {SIGMOD} Conference 2016, San Francisco,
  CA, USA, June 26 - July 01, 2016}}. \bibinfo{publisher}{{ACM}},
  \bibinfo{pages}{1363--1375}.
\newblock


\bibitem[\protect\citeauthoryear{Battle, Stonebraker, and Chang}{Battle
  et~al\mbox{.}}{2013}]%
        {battle2013dynamic}
\bibfield{author}{\bibinfo{person}{Leilani Battle}, \bibinfo{person}{Michael
  Stonebraker}, {and} \bibinfo{person}{Remco Chang}.}
  \bibinfo{year}{2013}\natexlab{}.
\newblock \showarticletitle{Dynamic reduction of query result sets for
  interactive visualizaton}. In \bibinfo{booktitle}{\emph{2013 IEEE
  International Conference on Big Data}}. IEEE, \bibinfo{pages}{1--8}.
\newblock


\bibitem[\protect\citeauthoryear{Bowman and Salem}{Bowman and Salem}{2007}]%
        {bowman2006semantic}
\bibfield{author}{\bibinfo{person}{Ivan~T Bowman} {and}
  \bibinfo{person}{Kenneth Salem}.} \bibinfo{year}{2007}\natexlab{}.
\newblock \showarticletitle{Semantic prefetching of correlated query
  sequences}. In \bibinfo{booktitle}{\emph{2007 IEEE 23rd International
  Conference on Data Engineering}}. IEEE, \bibinfo{pages}{1284--1288}.
\newblock


\bibitem[\protect\citeauthoryear{Chatzopoulou, Eirinaki, and
  Polyzotis}{Chatzopoulou et~al\mbox{.}}{2009}]%
        {chatzopoulou2009recom2}
\bibfield{author}{\bibinfo{person}{Gloria Chatzopoulou},
  \bibinfo{person}{Magdalini Eirinaki}, {and} \bibinfo{person}{Neoklis
  Polyzotis}.} \bibinfo{year}{2009}\natexlab{}.
\newblock \showarticletitle{Query recommendations for interactive database
  exploration}. In \bibinfo{booktitle}{\emph{International Conference on
  Scientific and Statistical Database Management}}. Springer,
  \bibinfo{pages}{3--18}.
\newblock


\bibitem[\protect\citeauthoryear{Chen, Zhang, Wu, Wang, and Xing}{Chen
  et~al\mbox{.}}{2021}]%
        {chen2021revisiting}
\bibfield{author}{\bibinfo{person}{Yu Chen}, \bibinfo{person}{Yong Zhang},
  \bibinfo{person}{Jiacheng Wu}, \bibinfo{person}{Jin Wang}, {and}
  \bibinfo{person}{Chunxiao Xing}.} \bibinfo{year}{2021}\natexlab{}.
\newblock \showarticletitle{Revisiting data prefetching for database systems
  with machine learning techniques}. In \bibinfo{booktitle}{\emph{2021 IEEE
  37th International Conference on Data Engineering (ICDE)}}. IEEE,
  \bibinfo{pages}{2165--2170}.
\newblock


\bibitem[\protect\citeauthoryear{Curino, Jones, Zhang, and Madden}{Curino
  et~al\mbox{.}}{2010}]%
        {curino2010schism}
\bibfield{author}{\bibinfo{person}{Carlo Curino}, \bibinfo{person}{Evan
  Philip~Charles Jones}, \bibinfo{person}{Yang Zhang}, {and}
  \bibinfo{person}{Samuel~R Madden}.} \bibinfo{year}{2010}\natexlab{}.
\newblock \showarticletitle{Schism: a workload-driven approach to database
  replication and partitioning}.
\newblock  (\bibinfo{year}{2010}).
\newblock


\bibitem[\protect\citeauthoryear{de~Zoysa, Bailey, and Borovica-Gajic}{de~Zoysa
  et~al\mbox{.}}{2025}]%
        {exploract}
\bibfield{author}{\bibinfo{person}{Kasun de Zoysa}, \bibinfo{person}{James
  Bailey}, {and} \bibinfo{person}{Renata Borovica-Gajic}.}
  \bibinfo{year}{2025}\natexlab{}.
\newblock \showarticletitle{ExplorAct: Context-Aware Next Action
  Recommendations for Interactive Data Exploration}. In
  \bibinfo{booktitle}{\emph{Proceedings of the 34th ACM International
  Conference on Information and Knowledge Management}}
  \emph{(\bibinfo{series}{CIKM '25})}. \bibinfo{publisher}{Association for
  Computing Machinery}, \bibinfo{address}{New York, NY, USA},
  \bibinfo{pages}{3746252.3761257}.
\newblock
\urldef\tempurl%
\url{https://doi.org/10.1145/3746252.3761257}
\showDOI{\tempurl}


\bibitem[\protect\citeauthoryear{Difallah, Pavlo, Curino, and
  Cudr{\'{e}}{-}Mauroux}{Difallah et~al\mbox{.}}{2013}]%
        {DifallahPCC13}
\bibfield{author}{\bibinfo{person}{Djellel~Eddine Difallah},
  \bibinfo{person}{Andrew Pavlo}, \bibinfo{person}{Carlo Curino}, {and}
  \bibinfo{person}{Philippe Cudr{\'{e}}{-}Mauroux}.}
  \bibinfo{year}{2013}\natexlab{}.
\newblock \showarticletitle{OLTP-Bench: An Extensible Testbed for Benchmarking
  Relational Databases}.
\newblock \bibinfo{journal}{\emph{Proceedings of the VLDB Endowment}}
  \bibinfo{volume}{7}, \bibinfo{number}{4} (\bibinfo{year}{2013}),
  \bibinfo{pages}{277--288}.
\newblock
\urldef\tempurl%
\url{https://doi.org/10.14778/2732240.2732246}
\showDOI{\tempurl}


\bibitem[\protect\citeauthoryear{Dimitriadou, Papaemmanouil, and
  Diao}{Dimitriadou et~al\mbox{.}}{2014}]%
        {dimitriadou2014explore}
\bibfield{author}{\bibinfo{person}{Kyriaki Dimitriadou}, \bibinfo{person}{Olga
  Papaemmanouil}, {and} \bibinfo{person}{Yanlei Diao}.}
  \bibinfo{year}{2014}\natexlab{}.
\newblock \showarticletitle{Explore-by-example: An automatic query steering
  framework for interactive data exploration}. In
  \bibinfo{booktitle}{\emph{Proceedings of the 2014 ACM SIGMOD international
  conference on Management of data}}. \bibinfo{pages}{517--528}.
\newblock


\bibitem[\protect\citeauthoryear{Dimitriadou, Papaemmanouil, and
  Diao}{Dimitriadou et~al\mbox{.}}{2016}]%
        {dimitriadou2016aide}
\bibfield{author}{\bibinfo{person}{Kyriaki Dimitriadou}, \bibinfo{person}{Olga
  Papaemmanouil}, {and} \bibinfo{person}{Yanlei Diao}.}
  \bibinfo{year}{2016}\natexlab{}.
\newblock \showarticletitle{AIDE: an active learning-based approach for
  interactive data exploration}.
\newblock \bibinfo{journal}{\emph{IEEE Transactions on Knowledge and Data
  Engineering}} \bibinfo{volume}{28}, \bibinfo{number}{11}
  (\bibinfo{year}{2016}), \bibinfo{pages}{2842--2856}.
\newblock


\bibitem[\protect\citeauthoryear{Ding, Chaudhuri, Gehrke, and Narasayya}{Ding
  et~al\mbox{.}}{2021}]%
        {dsb_drift}
\bibfield{author}{\bibinfo{person}{Bailu Ding}, \bibinfo{person}{Surajit
  Chaudhuri}, \bibinfo{person}{Johannes Gehrke}, {and} \bibinfo{person}{Vivek
  Narasayya}.} \bibinfo{year}{2021}\natexlab{}.
\newblock \showarticletitle{DSB: A decision support benchmark for
  workload-driven and traditional database systems}.
\newblock \bibinfo{journal}{\emph{Proceedings of the VLDB Endowment}}
  \bibinfo{volume}{14}, \bibinfo{number}{13} (\bibinfo{year}{2021}),
  \bibinfo{pages}{3376--3388}.
\newblock


\bibitem[\protect\citeauthoryear{Ding, Das, Marcus, Wu, Chaudhuri, and
  Narasayya}{Ding et~al\mbox{.}}{2019}]%
        {indexselect_AIMeetsAI}
\bibfield{author}{\bibinfo{person}{Bailu Ding}, \bibinfo{person}{Sudipto Das},
  \bibinfo{person}{Ryan Marcus}, \bibinfo{person}{Wentao Wu},
  \bibinfo{person}{Surajit Chaudhuri}, {and} \bibinfo{person}{Vivek~R.
  Narasayya}.} \bibinfo{year}{2019}\natexlab{}.
\newblock \showarticletitle{{AI} Meets {AI:} Leveraging Query Executions to
  Improve Index Recommendations}. In \bibinfo{booktitle}{\emph{Proceedings of
  the International Conference on Management of Data, {SIGMOD} Conference 2019,
  Amsterdam, The Netherlands, June 30 - July 5, 2019}}.
  \bibinfo{publisher}{{ACM}}, \bibinfo{pages}{1241--1258}.
\newblock
\urldef\tempurl%
\url{https://doi.org/10.1145/3299869.3324957}
\showDOI{\tempurl}


\bibitem[\protect\citeauthoryear{Doshi}{Doshi}{2003}]%
        {doshi2003adaptive}
\bibfield{author}{\bibinfo{person}{Punit~Rameshchandra Doshi}.}
  \bibinfo{year}{2003}\natexlab{}.
\newblock \emph{\bibinfo{title}{Adaptive prefetching for visual data
  exploration}}.
\newblock \bibinfo{thesistype}{Ph.D. Dissertation}. \bibinfo{school}{Worcester
  Polytechnic Institute}.
\newblock


\bibitem[\protect\citeauthoryear{Drosou and Pitoura}{Drosou and
  Pitoura}{2013}]%
        {drosou2013ymaldb}
\bibfield{author}{\bibinfo{person}{Marina Drosou} {and}
  \bibinfo{person}{Evaggelia Pitoura}.} \bibinfo{year}{2013}\natexlab{}.
\newblock \showarticletitle{Ymaldb: exploring relational databases via
  result-driven recommendations}.
\newblock \bibinfo{journal}{\emph{The VLDB Journal}} \bibinfo{volume}{22},
  \bibinfo{number}{6} (\bibinfo{year}{2013}), \bibinfo{pages}{849--874}.
\newblock


\bibitem[\protect\citeauthoryear{Du, Huang, and Sun}{Du et~al\mbox{.}}{2023}]%
        {du2023multi}
\bibfield{author}{\bibinfo{person}{Yihan Du}, \bibinfo{person}{Longbo Huang},
  {and} \bibinfo{person}{Wen Sun}.} \bibinfo{year}{2023}\natexlab{}.
\newblock \showarticletitle{Multi-task representation learning for pure
  exploration in linear bandits}. In \bibinfo{booktitle}{\emph{International
  Conference on Machine Learning}}. PMLR, \bibinfo{pages}{8511--8564}.
\newblock


\bibitem[\protect\citeauthoryear{Egger, Allali, Lissandrini, Mottin, and
  Karras}{Egger et~al\mbox{.}}{2026}]%
        {egger2026structural}
\bibfield{author}{\bibinfo{person}{Maximilian~K Egger}, \bibinfo{person}{Mehdi
  Allali}, \bibinfo{person}{Matteo Lissandrini}, \bibinfo{person}{Davide
  Mottin}, {and} \bibinfo{person}{Panagiotis Karras}.}
  \bibinfo{year}{2026}\natexlab{}.
\newblock \showarticletitle{Structural Normalization of Property Graphs}.
\newblock \bibinfo{journal}{\emph{Proceedings of the VLDB Endowment}}
  \bibinfo{volume}{19}, \bibinfo{number}{6} (\bibinfo{year}{2026}),
  \bibinfo{pages}{1400--1412}.
\newblock


\bibitem[\protect\citeauthoryear{Erling}{Erling}{2012}]%
        {erling2012virtuoso}
\bibfield{author}{\bibinfo{person}{Orri Erling}.}
  \bibinfo{year}{2012}\natexlab{}.
\newblock \showarticletitle{Virtuoso, a Hybrid RDBMS/Graph Column Store.}
\newblock \bibinfo{journal}{\emph{IEEE Data Eng. Bull.}} \bibinfo{volume}{35},
  \bibinfo{number}{1} (\bibinfo{year}{2012}), \bibinfo{pages}{3--8}.
\newblock


\bibitem[\protect\citeauthoryear{Fernandes, Bernardino,
  et~al\mbox{.}}{Fernandes et~al\mbox{.}}{2018}]%
        {fernandes2018graph}
\bibfield{author}{\bibinfo{person}{Diogo Fernandes}, \bibinfo{person}{Jorge
  Bernardino}, {et~al\mbox{.}}} \bibinfo{year}{2018}\natexlab{}.
\newblock \showarticletitle{Graph Databases Comparison: AllegroGraph, ArangoDB,
  InfiniteGraph, Neo4J, and OrientDB.}
\newblock \bibinfo{journal}{\emph{Data}}  \bibinfo{volume}{18}
  (\bibinfo{year}{2018}), \bibinfo{pages}{373--380}.
\newblock


\bibitem[\protect\citeauthoryear{G{\"o}{\c{c}}er, Tsakalidou, Nicholson, Kim,
  and Ailamaki}{G{\"o}{\c{c}}er et~al\mbox{.}}{2026}]%
        {goccer2026qvcache}
\bibfield{author}{\bibinfo{person}{An{\i}l~Eren G{\"o}{\c{c}}er},
  \bibinfo{person}{Ioanna Tsakalidou}, \bibinfo{person}{Hamish Nicholson},
  \bibinfo{person}{Kyoungmin Kim}, {and} \bibinfo{person}{Anastasia Ailamaki}.}
  \bibinfo{year}{2026}\natexlab{}.
\newblock \showarticletitle{QVCache: A Query-Aware Vector Cache}.
\newblock \bibinfo{journal}{\emph{arXiv preprint arXiv:2602.02057}}
  (\bibinfo{year}{2026}).
\newblock


\bibitem[\protect\citeauthoryear{Hu, Cai, Dinh, Xie, Yue, Chen, and Ooi}{Hu
  et~al\mbox{.}}{2025}]%
        {vectorDBsearch}
\bibfield{author}{\bibinfo{person}{Guoyu Hu}, \bibinfo{person}{Shaofeng Cai},
  \bibinfo{person}{Tien Tuan~Anh Dinh}, \bibinfo{person}{Zhongle Xie},
  \bibinfo{person}{Cong Yue}, \bibinfo{person}{Gang Chen}, {and}
  \bibinfo{person}{Beng~Chin Ooi}.} \bibinfo{year}{2025}\natexlab{}.
\newblock \showarticletitle{{HAKES:} Scalable Vector Database for Embedding
  Search Service}.
\newblock \bibinfo{journal}{\emph{Proc. {VLDB} Endow.}} \bibinfo{volume}{18},
  \bibinfo{number}{9} (\bibinfo{year}{2025}), \bibinfo{pages}{3049--3062}.
\newblock


\bibitem[\protect\citeauthoryear{Huang, Siddiqui, Alotaibi, Curino, Leeka,
  Jindal, Zhao, Camacho-Rodr{\'\i}guez, and Tian}{Huang et~al\mbox{.}}{2024}]%
        {huang2024sibyl}
\bibfield{author}{\bibinfo{person}{Hanxian Huang}, \bibinfo{person}{Tarique
  Siddiqui}, \bibinfo{person}{Rana Alotaibi}, \bibinfo{person}{Carlo Curino},
  \bibinfo{person}{Jyoti Leeka}, \bibinfo{person}{Alekh Jindal},
  \bibinfo{person}{Jishen Zhao}, \bibinfo{person}{Jes{\'u}s
  Camacho-Rodr{\'\i}guez}, {and} \bibinfo{person}{Yuanyuan Tian}.}
  \bibinfo{year}{2024}\natexlab{}.
\newblock \showarticletitle{Sibyl: Forecasting time-evolving query workloads}.
\newblock \bibinfo{journal}{\emph{Proceedings of the ACM on Management of
  Data}} \bibinfo{volume}{2}, \bibinfo{number}{1} (\bibinfo{year}{2024}),
  \bibinfo{pages}{1--27}.
\newblock


\bibitem[\protect\citeauthoryear{Idreos}{Idreos}{2013}]%
        {StratosExploration}
\bibfield{author}{\bibinfo{person}{Stratos Idreos}.}
  \bibinfo{year}{2013}\natexlab{}.
\newblock \bibinfo{booktitle}{\emph{Big Data Exploration}}.
\newblock \bibinfo{publisher}{Taylor and Francis}.
\newblock


\bibitem[\protect\citeauthoryear{Idreos, Papaemmanouil, and Chaudhuri}{Idreos
  et~al\mbox{.}}{2015}]%
        {idreos2015overview}
\bibfield{author}{\bibinfo{person}{Stratos Idreos}, \bibinfo{person}{Olga
  Papaemmanouil}, {and} \bibinfo{person}{Surajit Chaudhuri}.}
  \bibinfo{year}{2015}\natexlab{}.
\newblock \showarticletitle{Overview of data exploration techniques}. In
  \bibinfo{booktitle}{\emph{SIGMOD}}. \bibinfo{pages}{277--281}.
\newblock


\bibitem[\protect\citeauthoryear{Itzkovitz, Niv, and Schuster}{Itzkovitz
  et~al\mbox{.}}{2000}]%
        {dyn_repr_adapt}
\bibfield{author}{\bibinfo{person}{Ayal Itzkovitz}, \bibinfo{person}{Nitzan
  Niv}, {and} \bibinfo{person}{Assaf Schuster}.}
  \bibinfo{year}{2000}\natexlab{}.
\newblock \showarticletitle{Dynamic adaptation of sharing granularity in DSM
  systems}.
\newblock \bibinfo{journal}{\emph{Journal of Systems and Software}}
  \bibinfo{volume}{55}, \bibinfo{number}{1} (\bibinfo{year}{2000}),
  \bibinfo{pages}{19--32}.
\newblock


\bibitem[\protect\citeauthoryear{Jain, Moritz, Halperin, Howe, and
  Lazowska}{Jain et~al\mbox{.}}{2016}]%
        {sqlshare}
\bibfield{author}{\bibinfo{person}{Shrainik Jain}, \bibinfo{person}{Dominik
  Moritz}, \bibinfo{person}{Daniel Halperin}, \bibinfo{person}{Bill Howe},
  {and} \bibinfo{person}{Ed Lazowska}.} \bibinfo{year}{2016}\natexlab{}.
\newblock \showarticletitle{SQLShare: Results from a Multi-Year
  SQL-as-a-Service Experiment}. In \bibinfo{booktitle}{\emph{Proceedings of the
  International Conference on Management of Data, {SIGMOD} Conference 2016, San
  Francisco, CA, USA, June 26 - July 01, 2016}}. \bibinfo{publisher}{{ACM}},
  \bibinfo{pages}{281--293}.
\newblock
\urldef\tempurl%
\url{https://doi.org/10.1145/2882903.2882957}
\showDOI{\tempurl}


\bibitem[\protect\citeauthoryear{Kalinin, {\c{C}}etintemel, and Zdonik}{Kalinin
  et~al\mbox{.}}{2014}]%
        {kalinin2014interactive}
\bibfield{author}{\bibinfo{person}{Alexander Kalinin}, \bibinfo{person}{Ugur
  {\c{C}}etintemel}, {and} \bibinfo{person}{Stanley~B. Zdonik}.}
  \bibinfo{year}{2014}\natexlab{}.
\newblock \showarticletitle{Interactive data exploration using semantic
  windows}. In \bibinfo{booktitle}{\emph{International Conference on Management
  of Data, {SIGMOD} 2014, Snowbird, UT, USA, June 22-27, 2014}}.
  \bibinfo{publisher}{{ACM}}, \bibinfo{pages}{505--516}.
\newblock
\urldef\tempurl%
\url{https://doi.org/10.1145/2588555.2593666}
\showDOI{\tempurl}


\bibitem[\protect\citeauthoryear{Kersten, Idreos, Manegold, and Liarou}{Kersten
  et~al\mbox{.}}{2011}]%
        {ResearcherGuide}
\bibfield{author}{\bibinfo{person}{Martin~L. Kersten}, \bibinfo{person}{Stratos
  Idreos}, \bibinfo{person}{Stefan Manegold}, {and} \bibinfo{person}{Erietta
  Liarou}.} \bibinfo{year}{2011}\natexlab{}.
\newblock \showarticletitle{The Researcher's Guide to the Data Deluge: Querying
  a Scientific Database in Just a Few Seconds}.
\newblock \bibinfo{journal}{\emph{VLDB}} \bibinfo{volume}{4},
  \bibinfo{number}{12} (\bibinfo{year}{2011}), \bibinfo{pages}{1474--1477}.
\newblock


\bibitem[\protect\citeauthoryear{Khan, Segovia, and Kossmann}{Khan
  et~al\mbox{.}}{2018}]%
        {smartRouting}
\bibfield{author}{\bibinfo{person}{Arijit Khan}, \bibinfo{person}{Gustavo
  Segovia}, {and} \bibinfo{person}{Donald Kossmann}.}
  \bibinfo{year}{2018}\natexlab{}.
\newblock \showarticletitle{On smart query routing: for distributed graph
  querying with decoupled storage}. In \bibinfo{booktitle}{\emph{2018 USENIX
  Annual Technical Conference (USENIX ATC 18)}}. \bibinfo{pages}{401--412}.
\newblock


\bibitem[\protect\citeauthoryear{Kipf, Kipf, Radke, Leis, Boncz, and
  Kemper}{Kipf et~al\mbox{.}}{2018}]%
        {Learnedcardinality}
\bibfield{author}{\bibinfo{person}{Andreas Kipf}, \bibinfo{person}{Thomas
  Kipf}, \bibinfo{person}{Bernhard Radke}, \bibinfo{person}{Viktor Leis},
  \bibinfo{person}{Peter Boncz}, {and} \bibinfo{person}{Alfons Kemper}.}
  \bibinfo{year}{2018}\natexlab{}.
\newblock \showarticletitle{Learned cardinalities: Estimating correlated joins
  with deep learning}.
\newblock \bibinfo{journal}{\emph{arXiv preprint arXiv:1809.00677}}
  (\bibinfo{year}{2018}).
\newblock


\bibitem[\protect\citeauthoryear{Kumar, Quamar, Deshpande, and Khuller}{Kumar
  et~al\mbox{.}}{2014}]%
        {kumar2014sword}
\bibfield{author}{\bibinfo{person}{K~Ashwin Kumar}, \bibinfo{person}{Abdul
  Quamar}, \bibinfo{person}{Amol Deshpande}, {and} \bibinfo{person}{Samir
  Khuller}.} \bibinfo{year}{2014}\natexlab{}.
\newblock \showarticletitle{SWORD: workload-aware data placement and replica
  selection for cloud data management systems}.
\newblock \bibinfo{journal}{\emph{The VLDB Journal}} \bibinfo{volume}{23},
  \bibinfo{number}{6} (\bibinfo{year}{2014}), \bibinfo{pages}{845--870}.
\newblock


\bibitem[\protect\citeauthoryear{Lai, Zolaktaf, Milani, AlOmeir, Cao, and
  Pottinger}{Lai et~al\mbox{.}}{2023}]%
        {workloadQueryRecom}
\bibfield{author}{\bibinfo{person}{Eugenie~Yujing Lai}, \bibinfo{person}{Zainab
  Zolaktaf}, \bibinfo{person}{Mostafa Milani}, \bibinfo{person}{Omar AlOmeir},
  \bibinfo{person}{Jianhao Cao}, {and} \bibinfo{person}{Rachel Pottinger}.}
  \bibinfo{year}{2023}\natexlab{}.
\newblock \showarticletitle{Workload-Aware Query Recommendation Using Deep
  Learning.}. In \bibinfo{booktitle}{\emph{EDBT}}, Vol.~\bibinfo{volume}{23}.
  \bibinfo{pages}{53--65}.
\newblock


\bibitem[\protect\citeauthoryear{Lan, Bao, Culpepper, and Borovica-Gajic}{Lan
  et~al\mbox{.}}{2023}]%
        {lan2023learnedIndex}
\bibfield{author}{\bibinfo{person}{Hai Lan}, \bibinfo{person}{Zhifeng Bao},
  \bibinfo{person}{J~Shane Culpepper}, {and} \bibinfo{person}{Renata
  Borovica-Gajic}.} \bibinfo{year}{2023}\natexlab{}.
\newblock \showarticletitle{Updatable learned indexes meet disk-resident
  DBMS-from evaluations to design choices}.
\newblock \bibinfo{journal}{\emph{Proceedings of the ACM on Management of
  Data}} \bibinfo{volume}{1}, \bibinfo{number}{2} (\bibinfo{year}{2023}),
  \bibinfo{pages}{1--22}.
\newblock


\bibitem[\protect\citeauthoryear{Lan, Bao, Culpepper, Borovica-Gajic, and
  Dong}{Lan et~al\mbox{.}}{2024}]%
        {lan2024flearnedIndex}
\bibfield{author}{\bibinfo{person}{Hai Lan}, \bibinfo{person}{Zhifeng Bao},
  \bibinfo{person}{J~Shane Culpepper}, \bibinfo{person}{Renata Borovica-Gajic},
  {and} \bibinfo{person}{Yu Dong}.} \bibinfo{year}{2024}\natexlab{}.
\newblock \showarticletitle{A fully on-disk updatable learned index}. In
  \bibinfo{booktitle}{\emph{2024 IEEE 40th international conference on data
  engineering (ICDE)}}. IEEE, \bibinfo{pages}{4856--4869}.
\newblock


\bibitem[\protect\citeauthoryear{Li, Zhou, Li, and Gao}{Li
  et~al\mbox{.}}{2019}]%
        {li2019qtune}
\bibfield{author}{\bibinfo{person}{Guoliang Li}, \bibinfo{person}{Xuanhe Zhou},
  \bibinfo{person}{Shifu Li}, {and} \bibinfo{person}{Bo Gao}.}
  \bibinfo{year}{2019}\natexlab{}.
\newblock \showarticletitle{Qtune: A query-aware database tuning system with
  deep reinforcement learning}.
\newblock \bibinfo{journal}{\emph{Proceedings of the VLDB Endowment}}
  \bibinfo{volume}{12}, \bibinfo{number}{12} (\bibinfo{year}{2019}),
  \bibinfo{pages}{2118--2130}.
\newblock


\bibitem[\protect\citeauthoryear{Li, Li, Zhang, Borovica-Gajic, Wang, Zhang,
  Chen, Shi, Li, and Chen}{Li et~al\mbox{.}}{2025}]%
        {li2025agenttune}
\bibfield{author}{\bibinfo{person}{Yiyan Li}, \bibinfo{person}{Haoyang Li},
  \bibinfo{person}{Jing Zhang}, \bibinfo{person}{Renata Borovica-Gajic},
  \bibinfo{person}{Shuai Wang}, \bibinfo{person}{Tieying Zhang},
  \bibinfo{person}{Jianjun Chen}, \bibinfo{person}{Rui Shi},
  \bibinfo{person}{Cuiping Li}, {and} \bibinfo{person}{Hong Chen}.}
  \bibinfo{year}{2025}\natexlab{}.
\newblock \showarticletitle{Agenttune: An agent-based large language model
  framework for database knob tuning}.
\newblock \bibinfo{journal}{\emph{Proceedings of the ACM on Management of
  Data}} \bibinfo{volume}{3}, \bibinfo{number}{6} (\bibinfo{year}{2025}),
  \bibinfo{pages}{1--29}.
\newblock


\bibitem[\protect\citeauthoryear{Liu and Borovica-Gajic}{Liu and
  Borovica-Gajic}{2026}]%
        {drift}
\bibfield{author}{\bibinfo{person}{Guanli Liu} {and} \bibinfo{person}{Renata
  Borovica-Gajic}.} \bibinfo{year}{2026}\natexlab{}.
\newblock \showarticletitle{Toward Drift-Aware Database Benchmarking}.
\newblock \bibinfo{journal}{\emph{Proceedings of the VLDB Endowment}}
  \bibinfo{volume}{19}, \bibinfo{number}{8} (\bibinfo{year}{2026}),
  \bibinfo{pages}{1818--1825}.
\newblock
\urldef\tempurl%
\url{https://dl.acm.org/doi/abs/10.14778/3811243.3811254}
\showURL{%
\tempurl}


\bibitem[\protect\citeauthoryear{Ma, Ding, Han, and Zhang}{Ma
  et~al\mbox{.}}{2021}]%
        {ma2021metainsight}
\bibfield{author}{\bibinfo{person}{Pingchuan Ma}, \bibinfo{person}{Rui Ding},
  \bibinfo{person}{Shi Han}, {and} \bibinfo{person}{Dongmei Zhang}.}
  \bibinfo{year}{2021}\natexlab{}.
\newblock \showarticletitle{Metainsight: Automatic discovery of structured
  knowledge for exploratory data analysis}. In
  \bibinfo{booktitle}{\emph{Proceedings of the 2021 international conference on
  management of data}}. \bibinfo{pages}{1262--1274}.
\newblock


\bibitem[\protect\citeauthoryear{Marcus, Negi, Mao, Tatbul, Alizadeh, and
  Kraska}{Marcus et~al\mbox{.}}{2022}]%
        {qopt_bao}
\bibfield{author}{\bibinfo{person}{Ryan Marcus}, \bibinfo{person}{Parimarjan
  Negi}, \bibinfo{person}{Hongzi Mao}, \bibinfo{person}{Nesime Tatbul},
  \bibinfo{person}{Mohammad Alizadeh}, {and} \bibinfo{person}{Tim Kraska}.}
  \bibinfo{year}{2022}\natexlab{}.
\newblock \showarticletitle{Bao: Making Learned Query Optimization Practical}.
\newblock \bibinfo{journal}{\emph{{SIGMOD} Rec.}} \bibinfo{volume}{51},
  \bibinfo{number}{1} (\bibinfo{year}{2022}), \bibinfo{pages}{6--13}.
\newblock
\urldef\tempurl%
\url{https://doi.org/10.1145/3542700.3542703}
\showDOI{\tempurl}


\bibitem[\protect\citeauthoryear{Marcus, Negi, Mao, Zhang, Alizadeh, Kraska,
  Papaemmanouil, and Tatbul}{Marcus et~al\mbox{.}}{2019}]%
        {qopt_neo}
\bibfield{author}{\bibinfo{person}{Ryan Marcus}, \bibinfo{person}{Parimarjan
  Negi}, \bibinfo{person}{Hongzi Mao}, \bibinfo{person}{Chi Zhang},
  \bibinfo{person}{Mohammad Alizadeh}, \bibinfo{person}{Tim Kraska},
  \bibinfo{person}{Olga Papaemmanouil}, {and} \bibinfo{person}{Nesime Tatbul}.}
  \bibinfo{year}{2019}\natexlab{}.
\newblock \showarticletitle{Neo: {A} Learned Query Optimizer}.
\newblock \bibinfo{journal}{\emph{Proceedings of the VLDB Endowment}}
  \bibinfo{volume}{12}, \bibinfo{number}{11} (\bibinfo{year}{2019}),
  \bibinfo{pages}{1705--1718}.
\newblock
\urldef\tempurl%
\url{https://doi.org/10.14778/3342263.3342644}
\showDOI{\tempurl}


\bibitem[\protect\citeauthoryear{Mattson, Gecsei, Slutz, and Traiger}{Mattson
  et~al\mbox{.}}{1970}]%
        {LRU}
\bibfield{author}{\bibinfo{person}{R.~L. Mattson}, \bibinfo{person}{J. Gecsei},
  \bibinfo{person}{D.~R. Slutz}, {and} \bibinfo{person}{I.~L. Traiger}.}
  \bibinfo{year}{1970}\natexlab{}.
\newblock \showarticletitle{Evaluation Techniques for Storage Hierarchies}. In
  \bibinfo{booktitle}{\emph{IBM Systems Journal}}, Vol.~\bibinfo{volume}{9}.
  \bibinfo{pages}{78--117}.
\newblock


\bibitem[\protect\citeauthoryear{Megiddo and Modha}{Megiddo and Modha}{2003}]%
        {lru3_arc}
\bibfield{author}{\bibinfo{person}{Nimrod Megiddo} {and}
  \bibinfo{person}{Dharmendra~S Modha}.} \bibinfo{year}{2003}\natexlab{}.
\newblock \showarticletitle{$\{$ARC$\}$: A $\{$Self-Tuning$\}$, low overhead
  replacement cache}. In \bibinfo{booktitle}{\emph{2nd USENIX Conference on
  File and Storage Technologies (FAST 03)}}.
\newblock


\bibitem[\protect\citeauthoryear{Mortensen, Bainson, Tversted, Gr{\ae}m,
  Borovica-Gajic, Paudice, Mottin, and Karras}{Mortensen et~al\mbox{.}}{2026}]%
        {mortensen2026practical}
\bibfield{author}{\bibinfo{person}{Kasper~O Mortensen}, \bibinfo{person}{Ama~B
  Bainson}, \bibinfo{person}{Mathias~R Tversted}, \bibinfo{person}{Kristoffer~S
  Gr{\ae}m}, \bibinfo{person}{Renata Borovica-Gajic}, \bibinfo{person}{Andrea
  Paudice}, \bibinfo{person}{Davide Mottin}, {and} \bibinfo{person}{Panagiotis
  Karras}.} \bibinfo{year}{2026}\natexlab{}.
\newblock \showarticletitle{Practical Adversarial Multi-Armed Bandits with
  Sublinear Runtime}.
\newblock \bibinfo{journal}{\emph{Proceedings of Machine Learning and Systems}}
   \bibinfo{volume}{8} (\bibinfo{year}{2026}), \bibinfo{pages}{2143--2156}.
\newblock


\bibitem[\protect\citeauthoryear{Olma, Karpathiotakis, Alagiannis,
  Athanassoulis, and Ailamaki}{Olma et~al\mbox{.}}{2017}]%
        {olma2017slalom}
\bibfield{author}{\bibinfo{person}{Matthaios Olma}, \bibinfo{person}{Manos
  Karpathiotakis}, \bibinfo{person}{Ioannis Alagiannis}, \bibinfo{person}{Manos
  Athanassoulis}, {and} \bibinfo{person}{Anastasia Ailamaki}.}
  \bibinfo{year}{2017}\natexlab{}.
\newblock \showarticletitle{Slalom: Coasting through raw data via adaptive
  partitioning and indexing}.
\newblock \bibinfo{journal}{\emph{Proceedings of the VLDB Endowment}}
  \bibinfo{volume}{10}, \bibinfo{number}{10} (\bibinfo{year}{2017}),
  \bibinfo{pages}{1106--1117}.
\newblock


\bibitem[\protect\citeauthoryear{Opdenacker and Electrons}{Opdenacker and
  Electrons}{2007}]%
        {opdenacker2007readahead}
\bibfield{author}{\bibinfo{person}{Michael Opdenacker} {and}
  \bibinfo{person}{Free Electrons}.} \bibinfo{year}{2007}\natexlab{}.
\newblock \showarticletitle{Readahead: time-travel techniques for desktop and
  embedded systems}. In \bibinfo{booktitle}{\emph{Proc. of the 2007 Ottawa
  Linux Symposium}}, Vol.~\bibinfo{volume}{2}. \bibinfo{pages}{97--106}.
\newblock


\bibitem[\protect\citeauthoryear{Perera, Oetomo, Rubinstein, and
  Borovica-Gajic}{Perera et~al\mbox{.}}{2021}]%
        {dbabandit}
\bibfield{author}{\bibinfo{person}{R~Malinga Perera}, \bibinfo{person}{Bastian
  Oetomo}, \bibinfo{person}{Benjamin~IP Rubinstein}, {and}
  \bibinfo{person}{Renata Borovica-Gajic}.} \bibinfo{year}{2021}\natexlab{}.
\newblock \showarticletitle{DBA bandits: Self-driving index tuning under
  ad-hoc, analytical workloads with safety guarantees}. In
  \bibinfo{booktitle}{\emph{2021 IEEE 37th International Conference on Data
  Engineering (ICDE)}}. IEEE, \bibinfo{pages}{600--611}.
\newblock


\bibitem[\protect\citeauthoryear{Perera, Oetomo, Rubinstein, and
  Borovica-Gajic}{Perera et~al\mbox{.}}{2022}]%
        {hmab}
\bibfield{author}{\bibinfo{person}{R~Malinga Perera}, \bibinfo{person}{Bastian
  Oetomo}, \bibinfo{person}{Benjamin~IP Rubinstein}, {and}
  \bibinfo{person}{Renata Borovica-Gajic}.} \bibinfo{year}{2022}\natexlab{}.
\newblock \showarticletitle{HMAB: self-driving hierarchy of bandits for
  integrated physical database design tuning}.
\newblock \bibinfo{journal}{\emph{Proceedings of the VLDB}}
  \bibinfo{volume}{16}, \bibinfo{number}{2} (\bibinfo{year}{2022}),
  \bibinfo{pages}{216--229}.
\newblock


\bibitem[\protect\citeauthoryear{Perera, Oetomo, Rubinstein, and
  Borovica{-}Gajic}{Perera et~al\mbox{.}}{2023}]%
        {noDBA}
\bibfield{author}{\bibinfo{person}{R.~Malinga Perera}, \bibinfo{person}{Bastian
  Oetomo}, \bibinfo{person}{Benjamin I.~P. Rubinstein}, {and}
  \bibinfo{person}{Renata Borovica{-}Gajic}.} \bibinfo{year}{2023}\natexlab{}.
\newblock \showarticletitle{No DBA? No Regret! Multi-Armed Bandits for Index
  Tuning of Analytical and {HTAP} Workloads With Provable Guarantees}.
\newblock \bibinfo{journal}{\emph{{IEEE} Trans. Knowl. Data Eng.}}
  \bibinfo{volume}{35}, \bibinfo{number}{12} (\bibinfo{year}{2023}),
  \bibinfo{pages}{12855--12872}.
\newblock
\urldef\tempurl%
\url{https://doi.org/10.1109/TKDE.2023.3271664}
\showDOI{\tempurl}


\bibitem[\protect\citeauthoryear{Qin, Chen, and Zhu}{Qin et~al\mbox{.}}{2014}]%
        {qin2014contextual}
\bibfield{author}{\bibinfo{person}{Lijing Qin}, \bibinfo{person}{Shouyuan
  Chen}, {and} \bibinfo{person}{Xiaoyan Zhu}.} \bibinfo{year}{2014}\natexlab{}.
\newblock \showarticletitle{Contextual combinatorial bandit and its application
  on diversified online recommendation}. In
  \bibinfo{booktitle}{\emph{Proceedings of the 2014 SIAM International
  Conference on Data Mining}}. SIAM, \bibinfo{pages}{461--469}.
\newblock


\bibitem[\protect\citeauthoryear{Rodriguez, Yusuf, Lyons, Paz, Rangaswami, Liu,
  Zhao, and Narasimhan}{Rodriguez et~al\mbox{.}}{2021}]%
        {cacheus}
\bibfield{author}{\bibinfo{person}{Liana~V Rodriguez}, \bibinfo{person}{Farzana
  Yusuf}, \bibinfo{person}{Steven Lyons}, \bibinfo{person}{Eysler Paz},
  \bibinfo{person}{Raju Rangaswami}, \bibinfo{person}{Jason Liu},
  \bibinfo{person}{Ming Zhao}, {and} \bibinfo{person}{Giri Narasimhan}.}
  \bibinfo{year}{2021}\natexlab{}.
\newblock \showarticletitle{Learning cache replacement with $\{$CACHEUS$\}$}.
  In \bibinfo{booktitle}{\emph{19th USENIX Conference on File and Storage
  Technologies (FAST 21)}}. \bibinfo{pages}{341--354}.
\newblock


\bibitem[\protect\citeauthoryear{Sato and Ito}{Sato and Ito}{2025}]%
        {sato2025fast}
\bibfield{author}{\bibinfo{person}{Ryoma Sato} {and} \bibinfo{person}{Shinji
  Ito}.} \bibinfo{year}{2025}\natexlab{}.
\newblock \showarticletitle{Fast EXP3 Algorithms}.
\newblock \bibinfo{journal}{\emph{arXiv preprint arXiv:2512.11201}}
  (\bibinfo{year}{2025}).
\newblock


\bibitem[\protect\citeauthoryear{Schrott, Jakubowski, and Hose}{Schrott
  et~al\mbox{.}}{2027}]%
        {katjaEDBT27aNorm}
\bibfield{author}{\bibinfo{person}{Johannes Schrott}, \bibinfo{person}{Maxime
  Jakubowski}, {and} \bibinfo{person}{Katja Hose}.}
  \bibinfo{year}{2027}\natexlab{}.
\newblock \showarticletitle{A Graph-Native Approach to Normalization}. In
  \bibinfo{booktitle}{\emph{Proceedings 30th International Conference on
  Extending Database Technology, {EDBT} 2027, Lille, France, April 6-9, 2027}},
  \bibfield{editor}{\bibinfo{person}{Felix Naumann},
  \bibinfo{person}{Senjuti~Basu Roy}, \bibinfo{person}{Pierre Bourhis},
  \bibinfo{person}{Nofar Carmeli}, \bibinfo{person}{Fabian Panse}, {and}
  \bibinfo{person}{Suraj Shetiya}} (Eds.).
  \bibinfo{publisher}{OpenProceedings.org}, \bibinfo{pages}{61--73}.
\newblock
\urldef\tempurl%
\url{https://doi.org/10.48786/EDBT.2027.06}
\showDOI{\tempurl}


\bibitem[\protect\citeauthoryear{Serafini, Taft, Elmore, Pavlo, Aboulnaga, and
  Stonebraker}{Serafini et~al\mbox{.}}{2016}]%
        {2016clay}
\bibfield{author}{\bibinfo{person}{Marco Serafini}, \bibinfo{person}{Rebecca
  Taft}, \bibinfo{person}{Aaron~J Elmore}, \bibinfo{person}{Andrew Pavlo},
  \bibinfo{person}{Ashraf Aboulnaga}, {and} \bibinfo{person}{Michael
  Stonebraker}.} \bibinfo{year}{2016}\natexlab{}.
\newblock \showarticletitle{Clay: fine-grained adaptive partitioning for
  general database schemas}.
\newblock \bibinfo{journal}{\emph{Proceedings of the VLDB Endowment}}
  \bibinfo{volume}{10}, \bibinfo{number}{4} (\bibinfo{year}{2016}),
  \bibinfo{pages}{445--456}.
\newblock


\bibitem[\protect\citeauthoryear{Setiawan, Rubinstein, and
  Borovica{-}Gajic}{Setiawan et~al\mbox{.}}{2020}]%
        {interpolation}
\bibfield{author}{\bibinfo{person}{Naufal~Fikri Setiawan},
  \bibinfo{person}{Benjamin I.~P. Rubinstein}, {and} \bibinfo{person}{Renata
  Borovica{-}Gajic}.} \bibinfo{year}{2020}\natexlab{}.
\newblock \showarticletitle{Function Interpolation for Learned Index
  Structures}. In \bibinfo{booktitle}{\emph{Databases Theory and Applications -
  31st Australasian Database Conference, {ADC} 2020, Melbourne, VIC, Australia,
  February 3-7, 2020, Proceedings}} \emph{(\bibinfo{series}{Lecture Notes in
  Computer Science})}, \bibfield{editor}{\bibinfo{person}{Renata
  Borovica{-}Gajic}, \bibinfo{person}{Jianzhong Qi}, {and}
  \bibinfo{person}{Weiqing Wang}} (Eds.), Vol.~\bibinfo{volume}{12008}.
  \bibinfo{publisher}{Springer}, \bibinfo{pages}{68--80}.
\newblock
\urldef\tempurl%
\url{https://doi.org/10.1007/978-3-030-39469-1\_6}
\showDOI{\tempurl}


\bibitem[\protect\citeauthoryear{Smith}{Smith}{1978}]%
        {smith1978lookahead}
\bibfield{author}{\bibinfo{person}{Alan~Jay Smith}.}
  \bibinfo{year}{1978}\natexlab{}.
\newblock \showarticletitle{Sequentiality and prefetching in database systems}.
\newblock \bibinfo{journal}{\emph{ACM Transactions on Database Systems (TODS)}}
  \bibinfo{volume}{3}, \bibinfo{number}{3} (\bibinfo{year}{1978}),
  \bibinfo{pages}{223--247}.
\newblock


\bibitem[\protect\citeauthoryear{Song, Berger, Li, and Lloyd}{Song
  et~al\mbox{.}}{2020}]%
        {LRB}
\bibfield{author}{\bibinfo{person}{Zhenyu Song}, \bibinfo{person}{Daniel~S.
  Berger}, \bibinfo{person}{Kai Li}, {and} \bibinfo{person}{Wyatt Lloyd}.}
  \bibinfo{year}{2020}\natexlab{}.
\newblock \showarticletitle{Learning Relaxed Belady for Content Distribution
  Network Caching ({LRB})}. In \bibinfo{booktitle}{\emph{USENIX NSDI}}.
\newblock
\urldef\tempurl%
\url{https://www.cs.princeton.edu/~wlloyd/papers/lrb-nsdi20.pdf}
\showURL{%
\tempurl}


\bibitem[\protect\citeauthoryear{Song, Berger, Li, and Lloyd}{Song
  et~al\mbox{.}}{2023}]%
        {Halp}
\bibfield{author}{\bibinfo{person}{Zhenyu Song}, \bibinfo{person}{Daniel~S.
  Berger}, \bibinfo{person}{Kai Li}, {and} \bibinfo{person}{Wyatt Lloyd}.}
  \bibinfo{year}{2023}\natexlab{}.
\newblock \showarticletitle{{HALP}: Heuristic-Aided Learned Preferences for
  Cache Eviction}. In \bibinfo{booktitle}{\emph{USENIX NSDI}}.
\newblock
\urldef\tempurl%
\url{https://www.usenix.org/system/files/nsdi23-song-zhenyu.pdf}
\showURL{%
\tempurl}


\bibitem[\protect\citeauthoryear{Tauheed, Heinis, Sch{\"{u}}rmann, Markram, and
  Ailamaki}{Tauheed et~al\mbox{.}}{2012}]%
        {tauheed2012scout}
\bibfield{author}{\bibinfo{person}{Farhan Tauheed}, \bibinfo{person}{Thomas
  Heinis}, \bibinfo{person}{Felix Sch{\"{u}}rmann}, \bibinfo{person}{Henry
  Markram}, {and} \bibinfo{person}{Anastasia Ailamaki}.}
  \bibinfo{year}{2012}\natexlab{}.
\newblock \showarticletitle{{SCOUT:} Prefetching for Latent Feature Following
  Queries}.
\newblock \bibinfo{journal}{\emph{Proceedings of the VLDB Endowment}}
  \bibinfo{volume}{5}, \bibinfo{number}{11} (\bibinfo{year}{2012}),
  \bibinfo{pages}{1531--1542}.
\newblock


\bibitem[\protect\citeauthoryear{Wan, Garnett, and Ottley}{Wan
  et~al\mbox{.}}{2018}]%
        {wan2018learning}
\bibfield{author}{\bibinfo{person}{Ran Wan}, \bibinfo{person}{Roman Garnett},
  {and} \bibinfo{person}{Alvitta Ottley}.} \bibinfo{year}{2018}\natexlab{}.
\newblock \showarticletitle{Learning and Anticipating Future Actions During
  Exploratory Data Analysis}.
\newblock \bibinfo{journal}{\emph{arXiv preprint}} (\bibinfo{year}{2018}).
\newblock


\bibitem[\protect\citeauthoryear{Wang, Liu, Lin, Bao, Li, and Wang}{Wang
  et~al\mbox{.}}{2024}]%
        {indexAdvisor}
\bibfield{author}{\bibinfo{person}{Zijia Wang}, \bibinfo{person}{Haoran Liu},
  \bibinfo{person}{Chen Lin}, \bibinfo{person}{Zhifeng Bao},
  \bibinfo{person}{Guoliang Li}, {and} \bibinfo{person}{Tianqing Wang}.}
  \bibinfo{year}{2024}\natexlab{}.
\newblock \showarticletitle{Leveraging dynamic and heterogeneous workload
  knowledge to boost the performance of index advisors}.
\newblock \bibinfo{journal}{\emph{Proceedings of the VLDB Endowment}}
  \bibinfo{volume}{17}, \bibinfo{number}{7} (\bibinfo{year}{2024}),
  \bibinfo{pages}{1642--1654}.
\newblock


\bibitem[\protect\citeauthoryear{Wu and Ives}{Wu and Ives}{2024}]%
        {shifting_workload}
\bibfield{author}{\bibinfo{person}{Peizhi Wu} {and} \bibinfo{person}{Zachary~G
  Ives}.} \bibinfo{year}{2024}\natexlab{}.
\newblock \showarticletitle{Modeling shifting workloads for learned database
  systems}.
\newblock \bibinfo{journal}{\emph{Proceedings of the ACM on Management of
  Data}} \bibinfo{volume}{2}, \bibinfo{number}{1} (\bibinfo{year}{2024}),
  \bibinfo{pages}{1--27}.
\newblock


\bibitem[\protect\citeauthoryear{Wu, Negi, Alizadeh, Kraska, and Madden}{Wu
  et~al\mbox{.}}{2023}]%
        {wu2023factorjoin}
\bibfield{author}{\bibinfo{person}{Ziniu Wu}, \bibinfo{person}{Parimarjan
  Negi}, \bibinfo{person}{Mohammad Alizadeh}, \bibinfo{person}{Tim Kraska},
  {and} \bibinfo{person}{Samuel Madden}.} \bibinfo{year}{2023}\natexlab{}.
\newblock \showarticletitle{FactorJoin: a new cardinality estimation framework
  for join queries}.
\newblock \bibinfo{journal}{\emph{Proceedings of the ACM on Management of
  Data}} \bibinfo{volume}{1}, \bibinfo{number}{1} (\bibinfo{year}{2023}),
  \bibinfo{pages}{1--27}.
\newblock


\bibitem[\protect\citeauthoryear{Yang, Li, Shi, Li, Hu, Li, and jie Yuan}{Yang
  et~al\mbox{.}}{2023}]%
        {2023sgdp}
\bibfield{author}{\bibinfo{person}{Yiyuan Yang}, \bibinfo{person}{Rongshang
  Li}, \bibinfo{person}{Qiquan Shi}, \bibinfo{person}{Xijun Li},
  \bibinfo{person}{Gang Hu}, \bibinfo{person}{Xing Li}, {and}
  \bibinfo{person}{Min jie Yuan}.} \bibinfo{year}{2023}\natexlab{}.
\newblock \showarticletitle{SGDP: A Stream-Graph Neural Network Based Data
  Prefetcher}.
\newblock \bibinfo{journal}{\emph{2023 International Joint Conference on Neural
  Networks (IJCNN)}} (\bibinfo{year}{2023}), \bibinfo{pages}{1--8}.
\newblock


\bibitem[\protect\citeauthoryear{Zhang, Wu, Li, Tan, Li, and Cui}{Zhang
  et~al\mbox{.}}{2022}]%
        {onlineTune}
\bibfield{author}{\bibinfo{person}{Xinyi Zhang}, \bibinfo{person}{Hong Wu},
  \bibinfo{person}{Yang Li}, \bibinfo{person}{Jian Tan},
  \bibinfo{person}{Feifei Li}, {and} \bibinfo{person}{Bin Cui}.}
  \bibinfo{year}{2022}\natexlab{}.
\newblock \showarticletitle{Towards dynamic and safe configuration tuning for
  cloud databases}. In \bibinfo{booktitle}{\emph{Proceedings of the 2022
  International Conference on Management of Data}}. \bibinfo{pages}{631--645}.
\newblock


\bibitem[\protect\citeauthoryear{Zhang, Yue, Yang, Berger, Li, and Lloyd}{Zhang
  et~al\mbox{.}}{2023}]%
        {S3FIFO}
\bibfield{author}{\bibinfo{person}{Yazhuo Zhang}, \bibinfo{person}{Yao Yue},
  \bibinfo{person}{Jason Yang}, \bibinfo{person}{Daniel~S. Berger},
  \bibinfo{person}{Kai Li}, {and} \bibinfo{person}{Wyatt Lloyd}.}
  \bibinfo{year}{2023}\natexlab{}.
\newblock \showarticletitle{FIFO Queues are All You Need for Cache Eviction
  ({S3-FIFO})}. In \bibinfo{booktitle}{\emph{Proc. SOSP}}.
\newblock
\urldef\tempurl%
\url{https://jasony.me/publication/sosp23-s3fifo.pdf}
\showURL{%
\tempurl}


\bibitem[\protect\citeauthoryear{Zhou, Niu, Xiong, Fang, and Wang}{Zhou
  et~al\mbox{.}}{2025}]%
        {ThreeLCache}
\bibfield{author}{\bibinfo{person}{Wenbin Zhou}, \bibinfo{person}{Zhixiong
  Niu}, \bibinfo{person}{Yongqiang Xiong}, \bibinfo{person}{Juan Fang}, {and}
  \bibinfo{person}{Qian Wang}.} \bibinfo{year}{2025}\natexlab{}.
\newblock \showarticletitle{3L-Cache: Low Overhead and Precise Learning-based
  Eviction Policy for Caches}. In \bibinfo{booktitle}{\emph{USENIX FAST}}.
\newblock
\urldef\tempurl%
\url{https://www.usenix.org/system/files/fast25-zhou-wenbin.pdf}
\showURL{%
\tempurl}


\bibitem[\protect\citeauthoryear{Zirak, Choudhury, and Borovica{-}Gajic}{Zirak
  et~al\mbox{.}}{2024}]%
        {selep}
\bibfield{author}{\bibinfo{person}{Farzaneh Zirak}, \bibinfo{person}{Farhana
  Choudhury}, {and} \bibinfo{person}{Renata Borovica{-}Gajic}.}
  \bibinfo{year}{2024}\natexlab{}.
\newblock \showarticletitle{SeLeP: Learning Based Semantic Prefetching for
  Exploratory Database Workloads}.
\newblock \bibinfo{journal}{\emph{Proceedings of the VLDB Endowment}}
  \bibinfo{volume}{17}, \bibinfo{number}{8} (\bibinfo{year}{2024}),
  \bibinfo{pages}{2064--2076}.
\newblock


\bibitem[\protect\citeauthoryear{Zirak, Choudhury, and Borovica-Gajic}{Zirak
  et~al\mbox{.}}{2026}]%
        {grasp}
\bibfield{author}{\bibinfo{person}{Farzaneh Zirak}, \bibinfo{person}{Farhana
  Choudhury}, {and} \bibinfo{person}{Renata Borovica-Gajic}.}
  \bibinfo{year}{2026}\natexlab{}.
\newblock \showarticletitle{Generalizable Address-Aware Semantic Prefetching
  for Scalable Transactional and Analytical Workloads}. In
  \bibinfo{booktitle}{\emph{2026 IEEE 42nd International Conference on Data
  Engineering (ICDE)}}. IEEE, \bibinfo{pages}{1477--1490}.
\newblock


\end{thebibliography}

\end{document}